\documentclass[onecolumn]{emulateapj}

\usepackage{multirow}
\usepackage{graphicx}
\usepackage{amsmath, amsthm, amssymb}
\usepackage{color}
\usepackage{rotating}

\newcommand     \Msol   {M_{\odot}}

\newcommand{\beq}{\begin{equation}}
\newcommand{\eeq}{\end{equation}}
\newcommand{\beqa}{\begin{eqnarray}}
\newcommand{\eeqa}{\end{eqnarray}}

\countdef\decade=200
\advance\decade by \year
\countdef\hours=201
\advance\hours by \time
\divide\hours by 60
\countdef\mins=202
\advance\mins by \hours
\multiply\mins by 60
\multiply\hours by 100
\countdef\miltime=203
\advance\miltime by \hours
\advance\miltime by \time
\advance\miltime by -\mins

\begin{document}

\title{Kiloparsec-Scale Simulations of Star Formation in Disk Galaxies III. 
Structure and Dynamics of Filaments and Clumps in Giant Molecular Clouds}
\author{Michael J. Butler}
\affil{Institute for Computational Science, University of Zurich, 8049 Zurich, Switzerland}

\author{Jonathan C. Tan}
\affil{Departments of Astronomy \& Physics, University of Florida, Gainesville, FL 32611, USA}

\author{Sven Van Loo}
\affil{School of Physics and Astronomy, University of Leeds, Leeds LS2 9JT, UK}
\affil{Harvard-Smithsonian Center for Astrophysics, 60 Garden Street, Cambridge, MA 02138, USA}

\begin{abstract}

We present hydrodynamic simulations of self-gravitating dense gas in a
galactic disk, exploring scales ranging from 1 kpc down to $\sim
0.1$~pc.  Our primary goal is to understand how dense filaments form
in Giant Molecular Clouds (GMCs).  These structures, often observed as
Infrared Dark Clouds (IRDCs) in the Galactic plane, are thought to be
the precursors to massive stars and star clusters, so their formation
may be the rate limiting step controlling global star formation rates
in galactic systems as described by the Kennicutt-Schmidt
relation. Our study follows on from Van Loo et al. (2013, Paper I),
which carried out simulations to 0.5~pc resolution and examined global
aspects of the formation of dense gas clumps and the resulting star
formation rate. Here, using our higher resolution, we examine the
detailed structural, kinematic and dynamical properties of dense
filaments and clumps, including mass surface density ($\Sigma$)
probability distribution functions, filament mass per unit length and
its dispersion, lateral $\Sigma$ profiles, filament fragmentation,
filament velocity gradients and infall, and degree of filament and clump
virialization.
Where possible, these properties are compared to observations of
IRDCs.
By many metrics, especially too large mass fractions of high
$\Sigma>1\:{\rm g\:cm^{-2}}$ material, too high mass per unit length
dispersion due to dense clump formation, too high velocity gradients
and too high velocity dispersion for a given mass per unit length,
the simulated filaments differ from observed IRDCs. We thus conclude that
IRDCs do not form from global fast collapse of GMCs. Rather, we expect
IRDC formation and collapse is slowed significantly by the influence
of dynamically important magnetic fields, which may thus play a
crucial role in regulating galactic star formation rates.

\end{abstract}

\keywords{galaxies: ISM, galaxies: star clusters,
methods: numerical, ISM: structure, ISM: clouds, stars: formation}

\maketitle
\section{Introduction}

Stars form from molecular clouds in the interstellar medium (ISM),
especially giant molecular clouds (GMCs) (McKee \& Ostriker 2007
[MO07]; Tan et al. 2013). The rate of star formation appears to be
relatively inefficient compared to that derived from collapse of the
clouds at the free-fall rate (Zuckermann \& Evans 1974; Krumholz \&
Tan 2007). Part of the reason for this may be the high degree of
clustering of star formation within GMCs in regions with $A_V\gtrsim
10$~mag (Lada et al. 2010; Heiderman et al. 2010; Gutermuth et
al. 2011). Studying the formation of the dense substructures within
molecular clouds is thus crucial for a more complete understanding of
the star formation process.

Observationally, these dense substructures have been studied by:
various molecular line tracers, such as $\rm ^{13}CO$ (e.g., Jackson et
al. 2006), $\rm HCO^+$ (e.g., Barnes et al. 2011), and $\rm N_2H^+$
(e.g., Henshaw et al. 2013); sub-mm and mm dust continuum emission
(e.g., Rathborne et al. 2006; Ginsburg et al. 2012); and mid-infrared
extinction (e.g., Butler \& Tan 2009, 2012; Peretto \& Fuller 2009) of
Infrared Dark Clouds (IRDCs).

Theoretically, we expect gravitational collapse within molecular
clouds to be mediated by support from some combination of turbulence,
magnetic fields and stellar feedback (MO07). Some examples of recent
work studying dense gas formation include set-ups of internal GMC
converging flows (Chen \& Ostriker 2014), global turbulent clouds
(Smith et al. 2014), periodic box turbulence (Moeckel \& Burkert
2014), and formation of GMCs from converging atomic flows (Heitsch et
al. 2009; Gomez \& V\'azquez-Semadeni 2014).

Our approach differs from these previous studies by setting the
boundary conditions for GMCs from a galactic environment affected by
global galactic dynamics. Tasker \& Tan (2009, hereafter TT09) carried
out hydrodynamic simulations of an idealized axisymmetric, flat
rotation curve galaxy to resolve the formation and interaction of GMCs
(see also Dobbs 2008; Bournaud et al. 2010; Renaud et al. 2013). Their
mutual interactions lead to a supersonic velocity dispersion of the
clouds and frequent collisions that drive turbulence in the gas. In
order to understand the star formation process within molecular
clouds, a significant range of the cloud's internal structure then
needs to be resolved, ranging from the GMC-scale down to the $\sim
1$~pc scale clumps thought to represent the initial conditions of star
cluster formation. Van Loo et al. (2013, hereafter Paper I) followed a
1 kiloparsec-square patch of the TT09 disk (extended vertically for
$\pm 1$~kpc) down to a resolution of 0.5~pc for a period of
10~Myr. Star formation was introduced in gas above a threshold
``clump'' density of $n_{\rm H}=10^5\:{\rm cm^{-3}}$ and at a star
formation efficiency per local free-fall time of $\epsilon_{\rm
  ff}=0.02$. The star particles created, with a minimum mass of
100~$M_\odot$ representing clusters or subclusters of stars, were
prevented from accreting additional gas. Nor was local feedback from
these star particles implemented. In spite of the relatively low value
of $\epsilon_{\rm ff}$, the overall SFR seen in the simulation was
much, $\sim 100$ times, larger than in observed galaxies with similar
total mass surface densities of gas (e.g., Bigiel et al. 2008). This
was due in part to the much higher mass fractions of gas at ``GMC''
and ``clump'' densities than in real galaxies: about 70\% of the gas
was within ``GMCs'' (at $n_{\rm H}\geq 10^2\:{\rm cm^{-3}}$), and of
this about 50\% was also above the clump threshold density. In Paper
I, we speculated that inclusion of magnetic fields and/or local
feedback from young stars is needed to resolve this discrepancy.

In Van Loo, Tan \& Falle (2015, Paper II) we presented an initial
study of the effects of magnetic fields of various strengths on the
same kpc-scale patch of the galactic disk, finding modest levels of
global suppression of star formation rates by up to factors of
two. However, this result was strongly influenced by the presence of a
single magnetically supercritical starburst region in one part of the
simulation domain, and larger suppression factors were seen in other
regions.

Our goal in this paper, 
Paper III, is to follow the evolution of the GMCs and the
formation of dense filaments and clumps to higher spatial resolution,
$0.122$~pc. This is carried out with the same physics as modeled in
Paper I, namely pure hydrodynamics of self-gravitating gas (magnetic
fields 
at this resolution and stellar feedback are deferred to future papers).  The
rationale is to be able to carry out more detailed characterization of
the stucture, kinematics and dynamics of forming dense gas structures
for comparison with Galactic IRDCs, which, being at very early stages
of their star formation, are probably relatively unaffected by local
stellar feedback. Our particular focus is on the properties of long,
$\sim 50$~pc, filaments that form from the collapsing GMCs. We measure
various properties of simulated filaments and compare to similarly
long filaments, recently discovered, including as IRDCs, in the
Galactic interstellar medium (e.g., Jackson et al. 2010; Battersby \&
Bally 2012; Ragan et al. 2014).


These nonmagnetic, zero-feedback simulations should thus be regarded
as a baseline calculations from which we can then determine how much,
if any, suppression of collapse is needed from magnetic fields (and
stellar feedback) to more accurately represent the observed structure
and dynamics of dense, star-forming filaments and clumps.

In \S\ref{S:methods} we describe our methods and numerical set-up. In
\S\ref{S:resultsa} and \S\ref{S:resultsb} we describe our results. In \S\ref{S:conclusion} we
conclude.

\section{Methods and Numerical Set-up}\label{S:methods}

For our initial conditions, we use the same 1 kpc$^{2}$ patch ($x$-$y$
coordinates describe location in the disk plane) studied in Paper I,
that is centered at a galactocentric radius of 4.25~kpc (the galactic
center is 4.25~kpc away from the patch center in the negative $x$
direction and at the same $y$ value) and extending to $z=\pm1$~kpc,
above and below the disk. This patch was extracted from the global
galaxy simulation of TT09 after a time when the disk was largely
fragmented into a population of GMCs (see left column of
Figure~\ref{fig:boxsigma}). As in Paper I, the velocity field is
transformed to the frame of the circular velocity of 200~km~s$^{-1}$ at the
center of the box. Periodic boundary conditions are applied at the
$x-z$ faces of box and outflow boundary conditions at the other
faces. A fixed background potential is applied to represent the
vertical distribution of galactic stars and dark matter to match the
potential used by TT09. Note, this set-up is not that of a shearing
box (e.g., Coriolis forces are neglected), but the effects of this
approximation are expected to be quite small since the total time span
that is followed is relatively short compared to a flow crossing time
across the box.

Paper I followed the evolution of this region for 10~Myr with a
maximum resolution of 0.49~pc, but, to resolve structures on the
scale of individual star-forming clumps, higher resolution is
needed. The following simulations contain 6 levels of AMR on top of
the original 7.8~pc base grid resolution of the TT09 simulation. The
maximum resolution is then 0.122~pc, i.e., four times better than in
Paper I.

We include heating and cooling
    functions derived using the photodissociation code Cloudy (version 8; Ferland et al. 1998), as in Paper I.
These functions are able to treat gas at
temperatures as low as $5$~K (and up to $\sim 10^5$~K and beyond).
These functions include both atomic and molecular line cooling
processes, including
from H$_{2}$ and CO, among others.  A table of heating and cooling
rates for a range of densities and temperatures was generated based on
the density versus mean extinction relationship derived in Paper I.
For temperatures above $T = 10^{5}$~K, we opt to use the cooling curve
of Sarazin \& White (1987) and set the heating rate to zero.  For
densities and temperatures above or below the limits of the table, we
use the limiting rate.  For more details on the derivation of this
function, see Paper I.

Since our focus is on the dense, molecular gas, we adopt a fixed mean
particle mass of $\mu = 2.33 m_{\rm H} = 3.90\times 10^{-24}$~g, i.e.,
assuming $n_{\rm He}=0.1 n_{\rm H}$. Thus the sound speed is $c_{\rm
  th} = (\gamma P/\rho)^{1/2}$, which implies an adiabatic sound speed
$c_{\rm th} = (5 k T/[3\mu])^{1/2} \rightarrow 0.243 (T/10\:{\rm
  K})^{1/2}\:{\rm km\:s^{-1}}$. Our use of this fixed value of $\mu$
means that the pressures of the regions of our simulation that
correspond to atomic regions are smaller in our simulation by a factor
of 1.83 than they would be in reality.

We present the results of two separate simulation runs.  The first
run, Run nSF, includes all the above processes at a maximum resolution
of 0.122~pc.  In the second run, Run SF, we utilize a simple recipe
for star formation, which converts a fixed percentage of gas mass per
free-fall time into star particles if a cell exceeds a particular
threshold density.  We choose a value for the star formation
efficiency per local free-fall time, $\epsilon_{\rm ff}=0.02$
(Krumholz \& Tan 2007).  As in Paper I, we do not resolve individual
star-forming cores, so no requirements for the gas to be converging or
to be gravitationally bound are imposed.  When a cell exceeds the
threshold density, a star particle is formed whose mass is calculated
by
\begin{equation}\label{eq:mstar}
 M_* = \epsilon_{\rm ff} \frac{\rho \Delta x^3}{t_{\rm ff}} \Delta t,
\end{equation}
where $\rho$ is the gas density, $\Delta x^3$ the cell volume, $\Delta
t$ the numerical time step, and $t_{\rm ff}$ the free-fall time of gas
in the cell (evaluated as $t_{\rm ff} = (3\pi/32G\rho)^{1/2}$ with a
mean molecular weight of $\mu = 2.33$).  We use a threshold density of
$n_{\rm H} = 10^{6}$\ $\rm cm^{-3}$ and a minimum star particle mass
of 10~$\Msol$. If $M_* < M_{\rm min}$, then a particle is formed
stochastically with a probability $M_*/M_{\rm min}$.  At the threshold
density, the minimum gas mass in the cell is $M_{\rm *,min} =
63\ \Msol$. Note that the threshold density is higher and minimum star
particle mass smaller compared to Paper I, because of the higher grid
resolution.  The star particles are treated as collisionless particles
whose motions are governed by pure N-body calculations, and do not
gain any mass once they are formed.

The focus of this study is the collapse of GMCs to form filaments and
the fragmentation of these filaments to form dense star-forming
clumps.  Considering the time scales on which these processes occur,
we run the simulations for 4~Myr using the adaptive mesh refinement
hydrodynamics code {\it Enzo 2.0} (Bryan et al. 2014, O'Shea et
al. 2004).  After this time, the clouds have undergone significant
fragmentation and formed a large number of small pc-scale clumps in
several filaments (see Figure~\ref{fig:boxsigma}). In the run with
star formation a large number of star particles has been created. The
presence of a large number of dense gas and stellar structures 
makes continued calculation of the simulation very memory intensive,
slow and inefficient. Since continued evolution is not necessary for
us to achieve our scientific goals of studying the initial stages of
star formation and since other physics due star formation feedback is
not yet included in these models, we do not attempt to follow the
evolution beyond 4~Myr.

We note that the Zeus solver (Stone \& Norman 1992), rather than a
Godunov solver, has been used for these simulations. This introduces
relatively large heating rates due to numerical viscosity, but makes
the calculation more numerically stable. We have also carried out
simulation runs at the 0.5~pc resolution with this same solver for
comparison.


\section{Global Properties of the ISM and Star Formation}\label{S:resultsa}

\begin{figure}
\begin{center}$
\begin{array}{c} 
\includegraphics[width=6.5in]{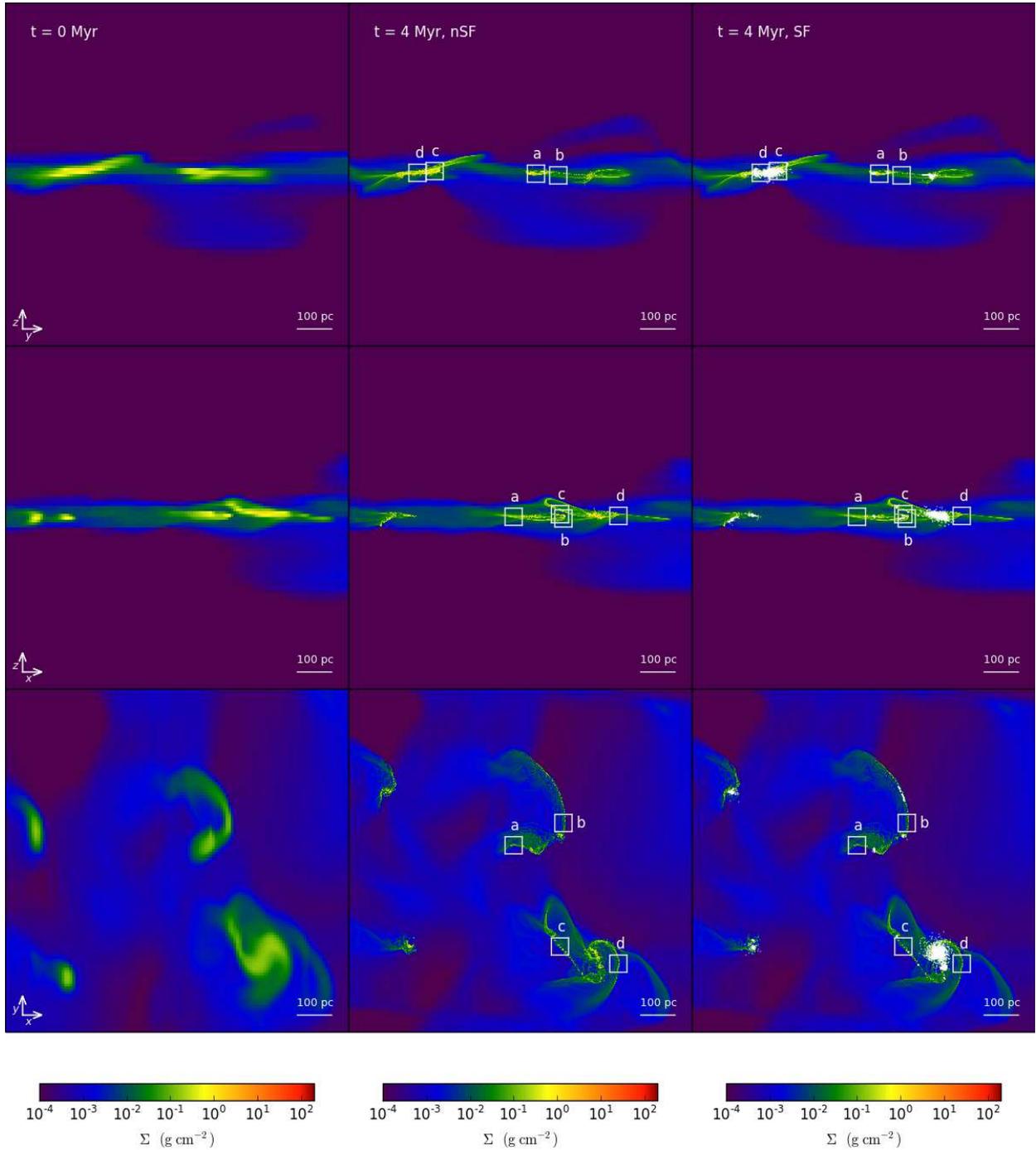} \\
\end{array}$
\end{center}
\caption{
Mass surface density, $\Sigma$, projections (in $\rm g\:cm^{-2}$)
along the $x$-axis (top row, equivalent to an in-galactic-plane view),
$y$-axis (middle row, equivalent to another in-galactic-plane view)
and $z$-axis (bottom row, equivalent to a top-down view of the
galactic plane) of the simulation box for the initial conditions (left
column), no star formation (nSF) run at 4.0 Myr (middle column), and
the star formation (SF) run at 4.0 Myr (right column; white dots
represent formed star particles). Also shown in the nSF and SF runs
are the locations of four 50~pc-cubed regions (a, b, c, d) around
filaments that have been selected for more detailed analysis (see
text).  We note that this figure is displayed at reduced resolution.
}\label{fig:boxsigma}
\end{figure}

\begin{figure}
\begin{center}$
\begin{array}{c} 
\includegraphics[width=6in]{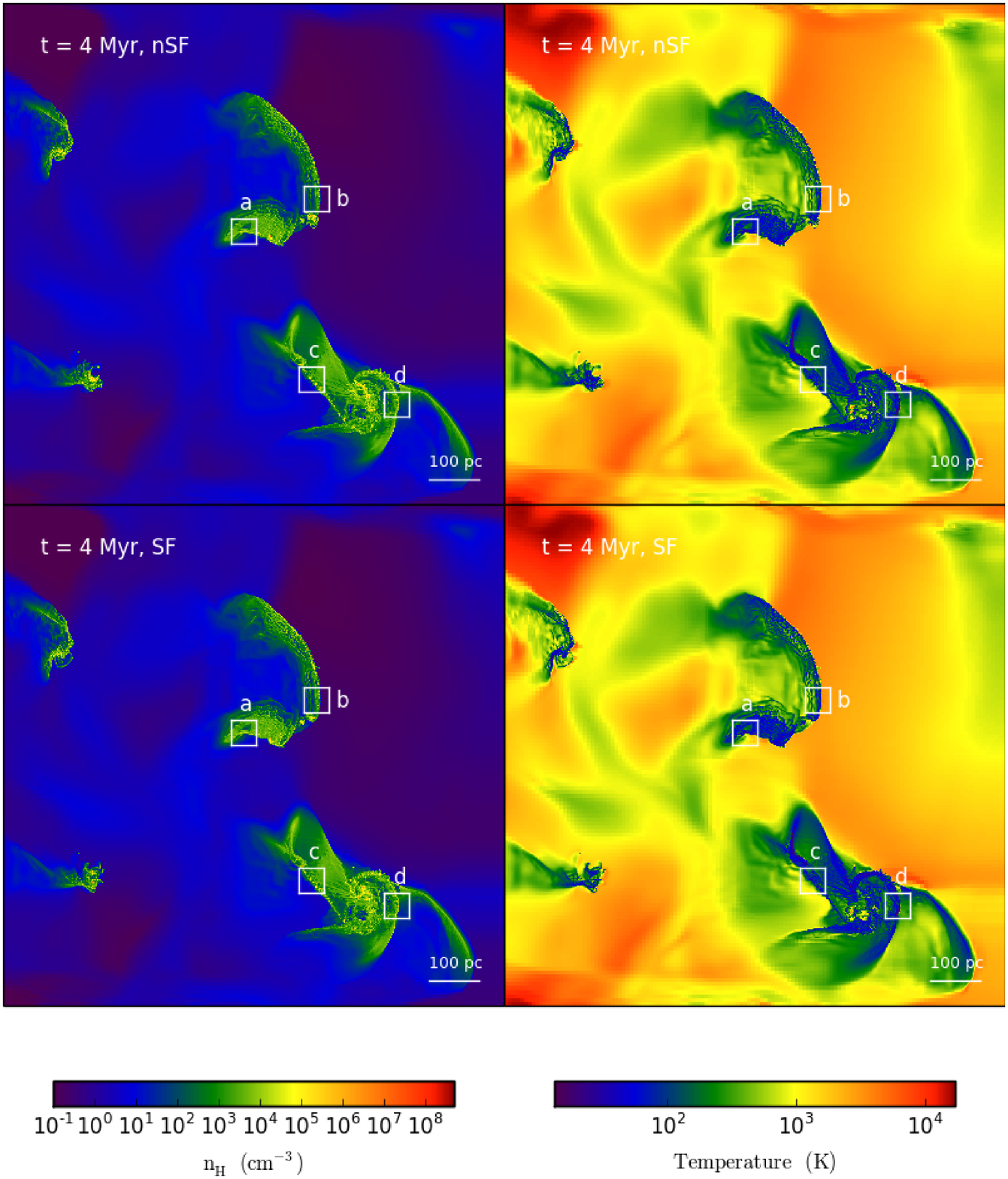}
\end{array}$
\end{center}
\caption{
Mass-weighted mean density, $n_{\rm H}$, (left column) and
temperature, $T$, (right column) of the top-down views of the nSF (top
row) and SF (bottom row) run simulations at 4~Myr.   We note that this figure is displayed at reduced resolution.
}\label{fig:boxtemp}
\end{figure}

Figure~\ref{fig:boxsigma} shows the mass surface density, $\Sigma$, of
the initial conditions (left column) and after 4 Myr of evolution of
Run nSF (middle column) and Run SF (right column) as seen along the
cardinal axes. The mean locations of the GMCs do not change much over
this relatively short timespan, given the clouds' initial
simulation-frame $x$-$y$-plane velocities of $\sim 20\:{\rm
  km\:s^{-1}}$ that are a mixture of the shearing velocity field of
galactic rotation and the peculiar motions imparted by gravitational
scattering interactions from prior evolution in the TT09
simulation. Note that these motions correspond to Mach numbers of
about 100 for gas that has been able to cool to $\sim 10$~K.

After 4~Myr, the clouds are seen to have collapsed to form filamentary
structures and more spheroidal clumps. The overall morphology of these
gas structures is very similar between the nSF and SF runs. The main
difference is seen in the peak $\Sigma$ values, with the SF run having
lower values due to conversion of gas into star particles.  The peak
$\Sigma$ values as seen in the top-down projections are 1250 and 317
$\rm g\:cm^{-2}$ in the nSF and SF runs, respectively.
Similarly, the peak gas density is $n_{\rm H} = 8.4 \times 10^{8}\: {\rm
cm}^{-3}$ and $n_{\rm H} = 4.1 \times 10^{8}\: {\rm cm}^{-3}$
in these runs. 

Figure~\ref{fig:boxtemp} shows the mass-averaged mean densities,
$n_{\rm H}$, (assuming $n_{\rm He}=0.1 n_{\rm H}$) and temperatures,
$T$, of the nSF and SF runs at 4~Myr. The GMC-like structures have
densities $\gtrsim 100\:{\rm cm^{-3}}$, and are often, but not always,
embedded in relatively dense HI structures with $n_{\rm H}\sim
10\:{\rm cm^{-3}}$ and temperatures of $\sim 100$--1000~K. There are
some cases, e.g., the GMC containing filament b, where the dense
molecular gas is moving into relatively low-density, warmer gas. We
note that since ionizing photon, stellar wind and supernova feedback
are not included, the global density and thermal structure of this
simulated ISM lacks the hot phase component and underrepresents the
warm phase components.

The total gas mass in Run nSF is $1.67 \times 10^{7}\: M_\odot$, while
it falls by $\sim 10\%$ to $1.50\times 10^{7}\: M_\odot$ in Run SF
after 4~Myr due to star formation activity. As a consequence the mass
in star particles grows to $1.67 \times 10^{6}\: M_\odot$, composed of
1.82 $\times 10^5$\ particles, resulting in a mean star particle mass
of 19.2~$M_\odot$. This means that, contrary to Paper I, not all star
particles are formed stochastically at the minimum mass level. In
fact, only a small percentage, 1.5\%, has the minimum star particle
mass of $10\:M_{\odot}$.  The maximum star particle mass in our
simulation is 289~$M_{\odot}$.
The average star formation rate per unit area during the 4.0~Myr of
the simulation is
$0.76\: M_{\odot}\: {\rm yr}^{-1}\: {\rm kpc}^{-2}$. 

\begin{figure}
\begin{center}$
\begin{array}{c} 
\includegraphics[width=6.5in]{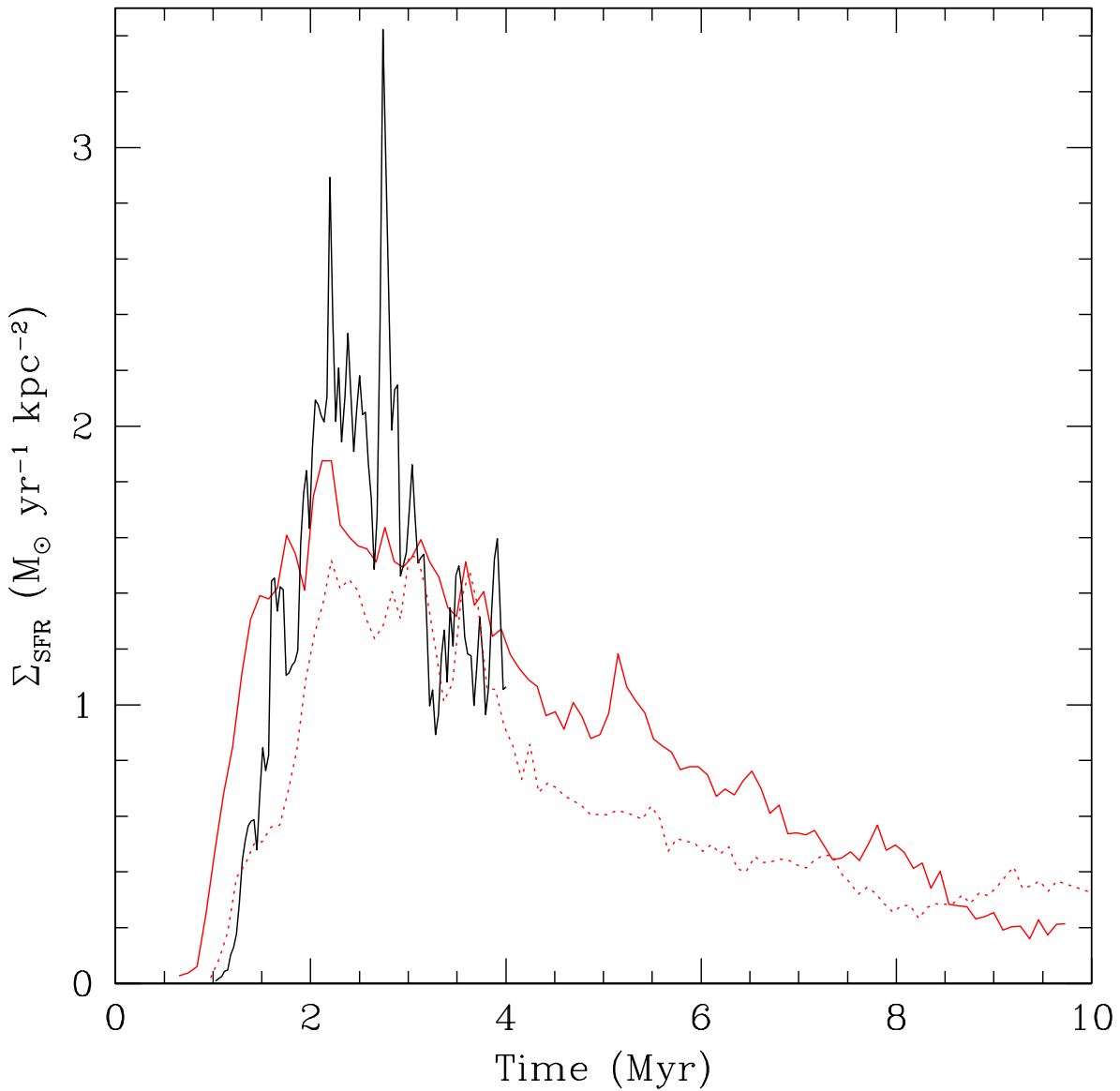} \\
 \end{array}$
\end{center}
\caption{
Time evolution of star formation rate per unit area for the
0.1~pc-resolution Run SF (black solid line) and the 0.5~pc-resolution
results from Paper I using the Godunov solver (red solid line),
and a 0.5~pc-resolution result using the Zeus solver (red dotted
line). 
%
}\label{fig:SFR}
\end{figure}

In Figure~\ref{fig:SFR}, we compare the SFR time evolution of Run SF
to those of the 0.5~pc resolution runs including star formation with a
density threshold of $n_{\rm H} = 10^{5}\: {\rm cm}^{-3}$ and a minimum
star particle mass of 100~$\Msol$ (equivalent to the simulations of
Paper I, but also investigating the effect of the use of the Zeus
rather than the Godunov solver). In all cases the evolution is
characterized by an onset of star formation after $\sim 1$~Myr, rising
to a peak at $\sim 2 - 3$~Myr, followed by a gradual decline. This
evolution should be viewed as a response to the initial conditions of
the simulation set-up, where dense, self-gravitating gas clouds are
released synchronously at $t=0$ and allowed to collapse to high
densities to form stars. After an initial burst of star formation, the
rate is seen to decline by factors of about 5 after 10~Myr. The choice
of hydrodynamics solver is seen to make a $\sim 10-20\%$ difference to
the SFR, with the additional heating introduced by the Zeus solver
reducing the SFR and delaying its onset and peak. The higher
resolution of Run SF, which involves a higher threshold density for
star formation, leads to increased and more time variable SFRs
compared to the lower resolution run that also uses the Zeus solver.

Similar to the results of Paper I, the overall SFRs per unit area are
much higher, by factors of $\sim 100$, than in observed galaxies
(e.g., Bigiel et al. 2008), most likely due to effects of magnetic
fields and/or feedback from newly formed stars that are present in
real galaxies, but lacking in these simulations. The quantitative
effect of magnetic fields for the same simulation set-up as Paper I is
being investigated by Van Loo et al. (in prep.).



%
%
%

%

\newpage
\section{Filament Structure, Kinematics and Dynamics}\label{S:resultsb}

\subsection{Filament Selection and Bulk Environmental Properties}

We select a sample of four large filaments, $a$ - $d$, from the
simulated GMCs at 4.0~Myr from the SF run for a detailed study of
their structure, kinematics and dynamics. These filaments are chosen
to sample a variety of environments and star formation
activities. Filaments a, b and d are still undergoing fragmentation
and collapse and have a lower level of star formation (it is expected
to increase as they continue to collapse), while Filament c is forming
stars actively. The average mass in these 50-pc-scale filament regions
is $\gtrsim 10^{5}\:M_{\odot}$ (see Table~\ref{tab:regions}). While
this is similar to the typical mass of a massive Galactic GMC, note
that these selected regions are just small parts of much larger and
more massive molecular clouds/complexes.

\begin{figure}
\begin{center}$
\begin{array}{c} 
\includegraphics[width=6.5in]{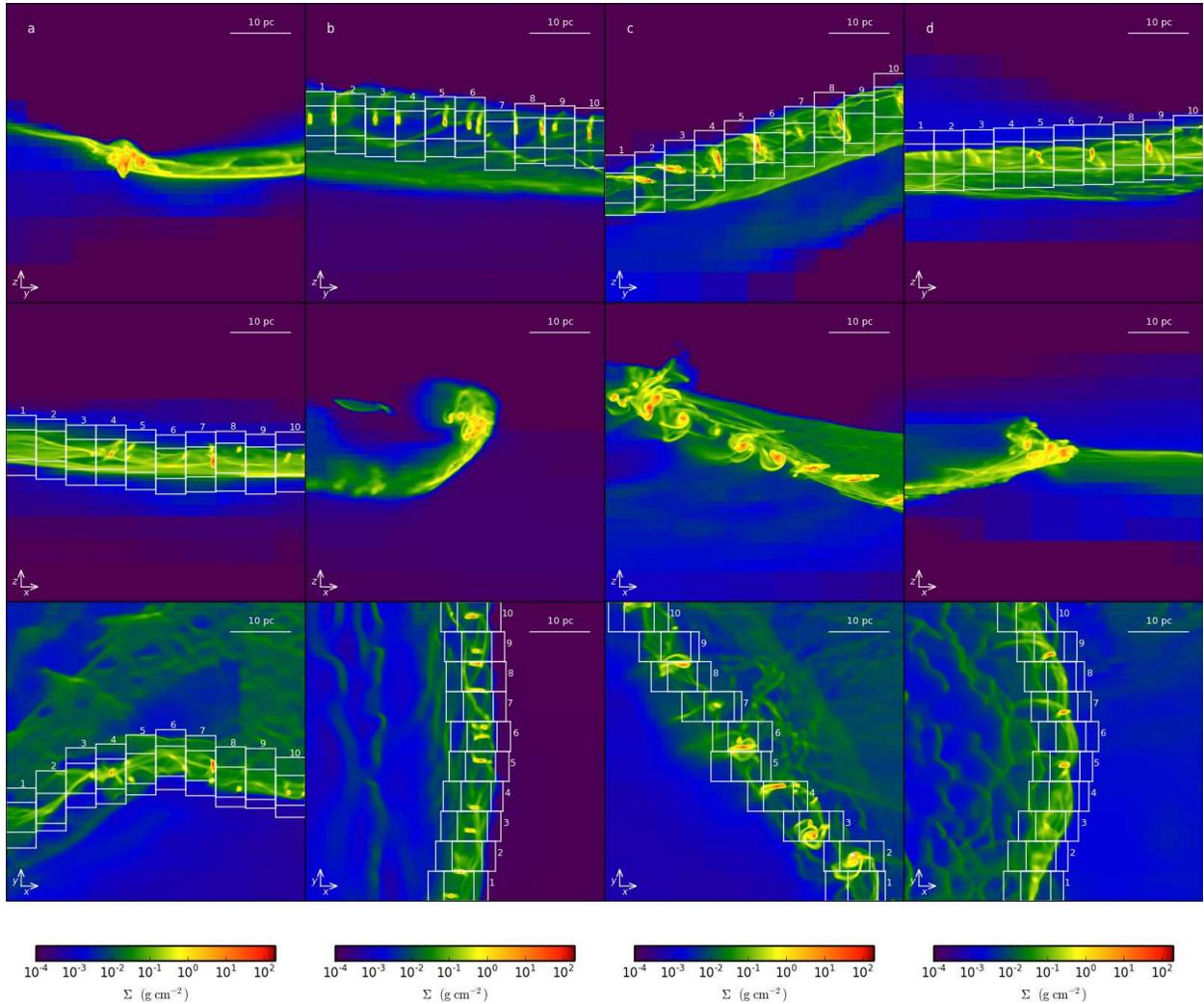} \\
\end{array}$
\end{center}
\caption{
Mass surface density, $\Sigma$, projections (in $\rm g\:cm^{-2}$)
along the (top to bottom) $x$, $y$, and $z$-axes centered on Filaments
a - d for the SF run at 4.0~Myr.
}\label{fig:filamentsigma}
\end{figure}

Mass surface density maps, with in-plane and top-down views, of the
50~pc-cubed regions containing the filaments are shown in
Figure~\ref{fig:filamentsigma}. The detailed structure of various
degrees of fragmentation in the filaments can be seen, along with the
structure of surrounding, more diffuse gas. The filaments tend to lie
in directions parallel to the galactic plane, with filament c showing
the largest deviation from an in-plane orientation.

Figure~\ref{fig:filamentsigma} also shows the division of the
filaments into ten ``strips'' (to be used for quantitative analysis of
filament properties, below), each of 5 pc width along the filament
(chosen to be $x$-direction for filament a; $y$-direction for b, c, d)
and 10 pc length perpendicular to the filament in each of the two
directions of this orthogonal plane. The central positions of the
strips in this plane are allowed to vary in order to track the
filament, with the position located by first centering on the center
of mass through the whole 50~pc region as viewed in the $\Sigma$
projections, and then re-centering on the center of mass within that
strip.


\begin{figure}
\begin{center}$
\begin{array}{cc} 
\includegraphics[width=6.5in]{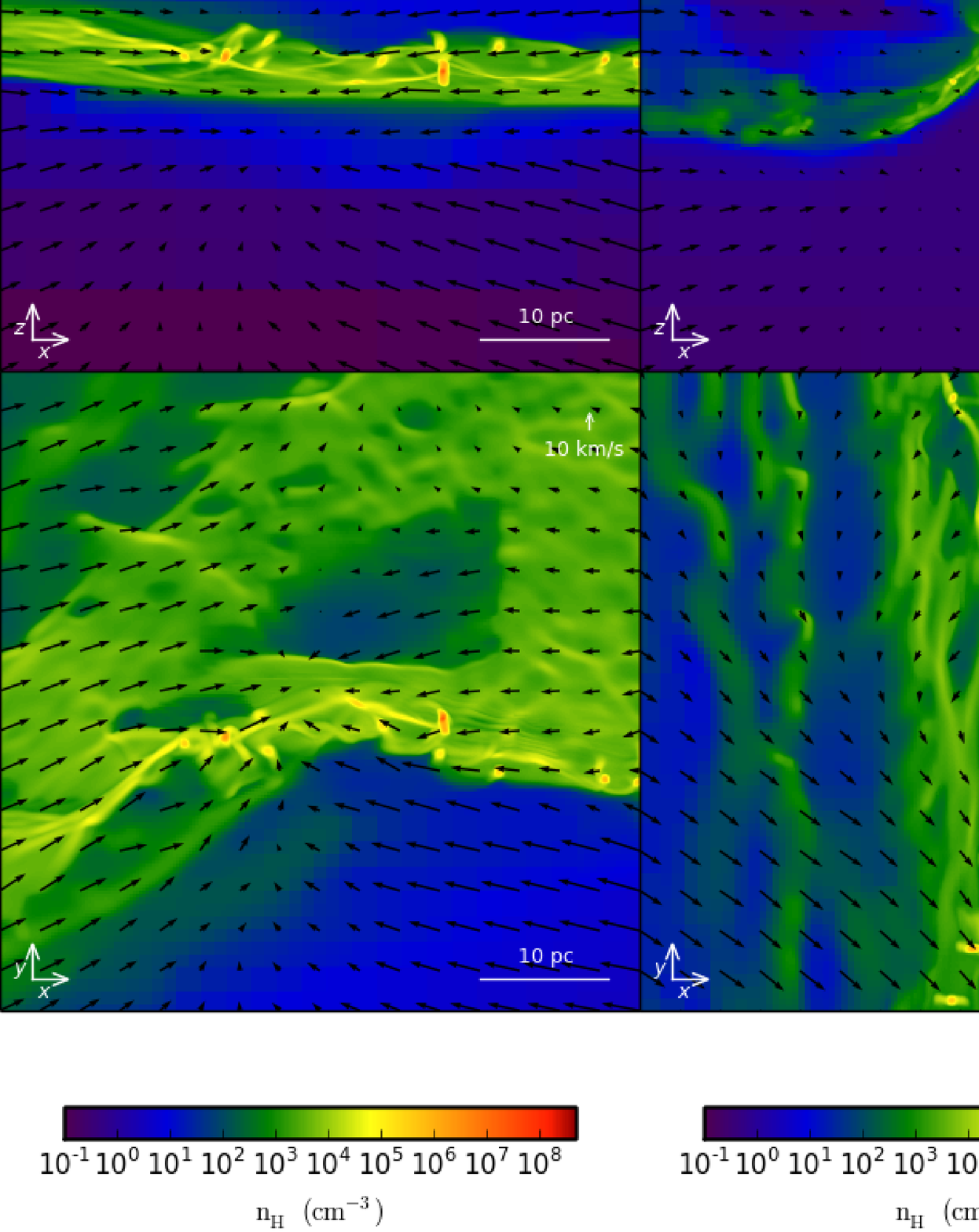} \\
\end{array}$
\end{center}
\caption{
Mass-weighted average densities, $n_{\rm H}$, along the (top to
bottom) $x$, $y$, and $z$-axes centered on Filaments $a$ - $d$ for the SF
run at 4.0~Myr.  Velocity vectors in the frame of the center of mass
of the region are overplotted in black.
}\label{fig:nHv}
\end{figure}

The volume densities, $n_{\rm H}$, mass-averaged through the 50 pc
filament regions, are shown in Figure~\ref{fig:nHv}. A wide range of
densities, from $\sim 10^{-1}$ to $\sim 10^7\:{\rm cm^{-3}}$ are
present. Note that, lacking wind, ionization and supernova feedback,
these simulations create this range of densities purely via
gravitational collapse, diffuse FUV heating and shocks resulting from
GMC motions and interactions.

The mass-average mean simulation-frame velocities of the material in
the 50~pc regions are typically $\sim 20$~km~s$^{-1}$ (see
Table~\ref{tab:regions}). The local velocity fields with respect to
these region velocities are also shown in Figure~\ref{fig:nHv}, with
typical values of $\sim 10$~km~s$^{-1}$. Some large-scale converging flows
are seen around the dense gas structures, together with other more
disordered flows.

\begin{figure}
\begin{center}$
\begin{array}{cc} 
\includegraphics[width=6.5in]{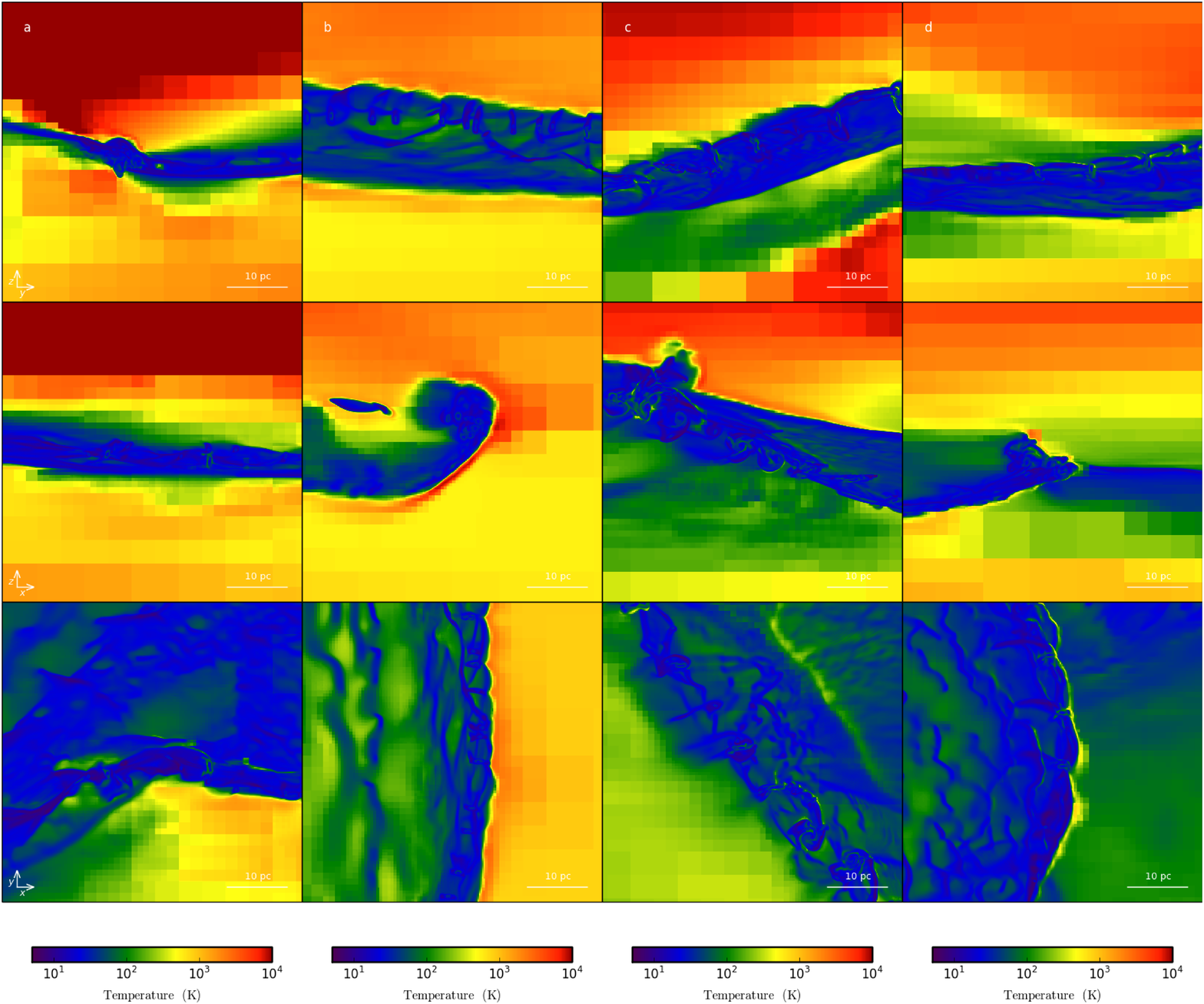} \\
\end{array}$
\end{center}
\caption{
Mass-weighted average temperature, $T$, along the (top to bottom) $x$,
$y$, and $z$-axes centered on Filaments a--d for the SF run at
4.0~Myr. Note, that all temperatures above $10^4$~K are indicated with
the same color, to enable greater diagnostic power in the range from 5
to $10^4$~K, but some shock-heated hotter components are present up to
$\sim 5 \times 10^6$~K (see text).
}\label{fig:temp}
\end{figure}

The mass-weighted temperatures of the gas along the various sight
lines of the regions are shown in Figure~\ref{fig:temp}. Again a wide
range of values are present, with the densest gas able to cool to
$\lesssim 10$~K and low density regions reaching $\sim 10^4$~K from
heating from the diffuse FUV radiation field and much higher
temperatures ($\sim 5 \times 10^{6}$~K) from shocks, with speeds of
$\sim$10 to 30~km~s$^{-1}$.

The probability distribution function (PDF) of mass surface density
(or equivalently column density) can provide useful insight into the
processes governing a cloud's structure.  Kainulainen et al. (2009)
showed that clouds undergoing star formation have wider column density
PDFs, including a power law tail of high column densities, while
non-star-forming clouds have narrower distributions, better fit by a
single log-normal function.  The formation of PDF power law tails has
been interpreted as being due to evolution of a separate
self-gravitating component of the clouds that are undergoing free-fall
collapse (e.g., Kritsuk et al. 2011; Ballesteros-Paredes et al. 2011).
In this case the development of such PDF tails would mark the onset of
star formation.

\begin{figure}
\begin{center}$
\begin{array}{cc} 
\includegraphics[width=6.0in]{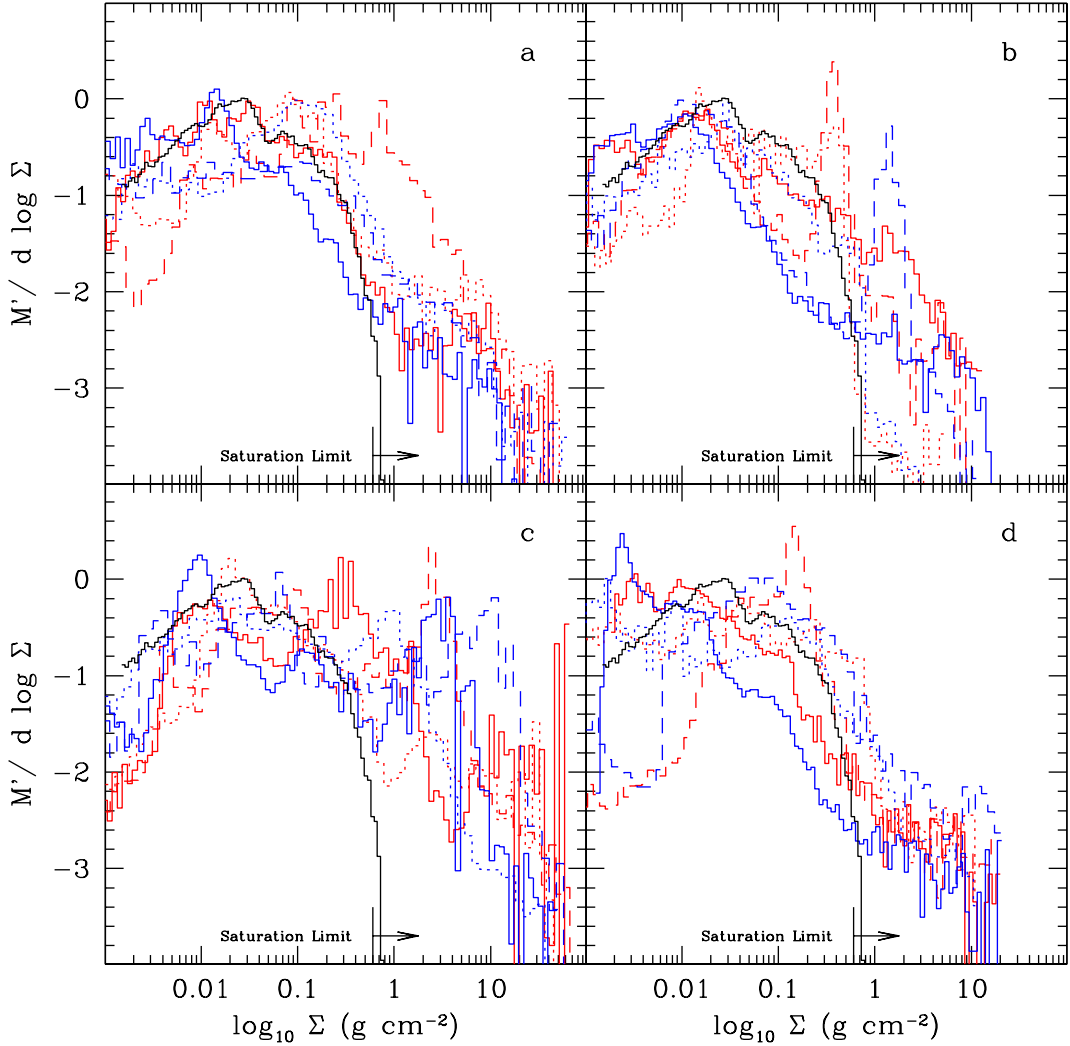}
\end{array}$
\end{center}
\caption{
Mass-weighted column density PDFs for Filament $a$ (top left),
Filament $b$ (top right), Filament $c$ (bottom left), Filament $d$
(bottom right).  Mass surface density PDFs of the inner 25 x 25 x 25
pc cube (red) and 50 x 50 x 50 pc cube (blue) calculated from the
density projection along the z-axis are shown as solid lines, along
the y-axis as dotted lines, and along the x-axis as dashed lines.
Also plotted is the MIR+NIR extinction mapping PDF for Cloud C (BTK13,
black solid line).  The region affected by the saturation limit, i.e.,
where $\Sigma > \Sigma_{\rm sat}=0.6\:{\rm g\:cm^{-2}}$, is indicated. 
}\label{fig:sigpdf}
\end{figure}

For each of the four filament regions in the SF run, we construct the
mass-weighted $\Sigma$ PDF for the 50-pc cube as viewed in each of the
three orthogonal axes, $x$, $y$, $z$, and these are presented as the
blue lines in Figure~\ref{fig:sigpdf}. A fair amount of structure can
be present in the PDFs at all $\Sigma$ values, which we expect is
partly a consequence of the global cloud structure that happens to be
within the defined region. At the high $\Sigma$ end of the
distributions, power-law tails are often present, but can show
variation in their properties depending on the viewing angle.

We also zoom into a 25-pc cube, centered on the center of mass of the
regions, and show the $\Sigma$ PDFs of these regions (red lines in
Figure~\ref{fig:sigpdf}). The differences compared to the 50-pc cube
regions illustrate the effect that the boundary region definition has
on the PDF, including the normalization level of the power law tails,
which rise as one zooms in on the denser region.

Observationally, $\Sigma$ PDFs can be measured in a number of ways,
including via sub-mm dust emission (which requires also knowing the
temperature structure of the clouds and dust emissivity properties per
unit total mass) and via NIR or MIR extinction (which requires knowing
dust opacities per unit total mass, but not cloud temperatures). In
Figure~\ref{fig:sigpdf} we show the $\Sigma$~PDF of IRDC C from the
BT09 sample (G028.37+00.07) as measured in the study of Butler et
al. (2014). This is one of the most massive and highest column density
IRDCs known in the Galaxy. The PDF has been measured in a region that
is designed to be complete for $A_V\gtrsim 3$~mag (i.e.,
$\Sigma\gtrsim 0.015\:{\rm g\:cm^{-2}}$), extending about
20\arcmin\ on a side (i.e., about 29~pc at the IRDC distance of
5~kpc). Note that the MIR extinction mapping technique has an upper
limit of $\Sigma$ that it is able to probe, corresponding to $\sim
0.6\:{\rm g\:cm^{-2}}$. This region can be probed by sub-mm emission
studies (e.g., Battersby et al. 2011; Schneider et al. 2014), 
but these results are dependent on the accuracy of the derived dust
temperature and assumed emissivity properties, and they also have
lower spatial resolution compared to the MIR extinction maps.



Allowing for the inability of the MIR extinction map to probe to very
high $\Sigma$ values, the overall comparison is quite favorable.
However, it should be noted that IRDC C is one of the more extreme
examples known, while the filament regions selected in the simulation
are quite moderate examples of dense regions. More typical IRDCs,
selected from the BT12 sample, have been studied by Kainulainen \& Tan
(2013), and these have relatively smaller amounts of gas at higher
$\Sigma$ values. The GMCs and nearby star-forming regions studied by
Kainulainen et al. (2009) show even smaller high $\Sigma$ mass
fractions. The largest values of $\Sigma$ that are seen within the
kpc-scale volume of the simulations are $>100\:{\rm g\:cm^{-2}}$, much
higher than have been seen via mm dust emission in IRDCs or
star-forming clouds.

We evaluate the area and mass fractions of the PDFs that are
$>0.1\:{\rm g\:cm^{-2}}$ and $>1\:{\rm g\:cm^{-2}}$ and list the
results in Table~\ref{tab:regions}. There can be a large variation
depending on orientation, i.e., if the viewing direction is
perpendicular or parallel to the main axis of the filament. Again
these values appear to be relatively high compared to observed IRDCs
(e.g., KT13, BTK14), although care must be taken to account for the
completeness limit of the observed $\Sigma$ distribution.

Still, given that both saturated and IR-bright regions of IRDCs
(where the extinction mapping method fails) cover very small
fractional areas (Butler \& Tan 2012), we do not expect these
limitations of the measured $\Sigma$ PDFs to change the overall
conclusion that the simulated clouds have much higher mass fractions
at high $\Sigma$.

\begin{sidewaystable*}
\setlength\arraycolsep{3pt}
\tiny
\caption{Filament 25 and 50-pc Region Properties$^a$}

\setlength{\tabcolsep}{0.035in} 
\begin{tabular}{c|c|c|c|c|c|c|c}


\tableline
\tableline
Filament & $x_{c}$,$y_{c}$,$z_{c}$ & $M_g$ & $n_{\rm H}$ & $f_{\Sigma0.1,x}$,$f_{\Sigma0.1,y}$,$f_{\Sigma0.1,z}$  & $f_{\Sigma1,x}$,$f_{\Sigma1,y}$,$f_{\Sigma1,z}$  &   $v_{x}$,$v_{y}$,$v_{z}$ & $\sigma_{v,x}$,$\sigma_{v,y}$,$\sigma_{v,z}$ \\
         &                       &                                           & $\bar{n}_{\rm H}$   & $\bar{f}_{\Sigma0.1,x}$,$\bar{f}_{\Sigma0.1,y}$,$\bar{f}_{\Sigma0.1,z}$ & $\bar{f}_{\Sigma1,x}$,$\bar{f}_{\Sigma1,y}$,$\bar{f}_{\Sigma1,z}$  & $\bar{v}_{x}$,$\bar{v}_{y}$,$\bar{v}_{z}$ & $\bar{\sigma}_{v,x}$,$\bar{\sigma}_{v,y}$,$\bar{\sigma}_{v,z}$ \\
         & (pc)                  & (10$^4$ $M_\odot$)     & ($10^{4}$ cm$^{-3}$) & ($10^{-2}$) & ($10^{-2}$) & (km\ s$^{-1}$)  & (km\ s$^{-1}$)  \\

\tableline
$a$ &   480[482],545[546],3.00[1.38]       & 22.3[12.4]  & 0.00517[0.0124]  & 3.81[4.93],4.97[6.05],1.39[3.49]   & 0.264[0.763],0.0880[0.294],0.0794[0.257]   & -22.6[-24.4],2.94[3.30],-1.81[-1.74] & 5.37[5.09],6.54[7.86],3.18[3.90]  \\
    &                                        &          & 145[259]         & 10.9[67.3],37.9[22.3],3.04[13.3]   & 0.800[7.32],0.746[1.16],0.271[0.424]      & -20.7[-20.9],7.62[9.18],-1.87[-2.00] & 5.75[5.29],7.20[8.41],5.62[7.39] \\
\tableline                                                         
$b$ &   625[626],610[612],-3.50[-1.05] & 14.2[5.20]     & 0.00329[0.00964] & 0.873[1.99],1.93[4.04],0.939[2.00] & 0.140[0.245],0.211[0.294],0.144[0.266]  & -9.05[-8.49],-10.3[-9.99],-1.01[-1.12]   & 5.93[5.63],3.85[2.60],2.58[2.46]  \\
    &                                      &            & 42.7[13.1]       & 7.97[36.3],5.76[21.2],0.985[9.02]  & 6.90[0.357],0.0142[0.0151],0.258[1.49] & -7.44[-6.56],-9.97[-9.89],-0.561[-0.519] & 4.53[3.35],4.31[2.50],3.61[2.65]   \\
\tableline                                               
$c^b$ &   614[618],251[248],20.3[19.1]       & 38.6[17.3]  &  0.00895[0.0321] & 2.37[4.03],3.72[7.31],2.53[4.92]  & 0.409[0.846],0.470[1.20],0.420[0.864]   & -20.4[-18.9],-11.6[-12.3],-0.433[-0.370] & 7.35[7.78],6.21[5.63],4.37[4.17]  \\
    &                                      &            &  326[268]        & 38.9[48.2],30.4[12.8],27.7[56.0]  & 28.7[35.7],7.89[2.02],19.8[7.73]        & -13.9[-13.5],-7.69[-8.81],-1.17[-1.35]   & 8.90[9.19],6.80[4.50],6.48[8.07]  \\
\tableline
$d$ &   785[783],200[202],10.0[5.98]    & 20.0[8.63]    & 0.00464[0.0160]     & 3.53[4.37],2.93[3.99],1.95[3.26]  &  0.111[0.169],0.262[0.393],0.125[0.226] & -22.7[-21.4],7.17[7.74],1.23[1.16]         & 7.50[7.48],8.87[8.00],2.85[2.83]  \\
    &                                   &               & 44.9[59.6]          & 29.7[62.4],16.3[19.8],1.26[4.55]  &  0.948[0.264],0.321[0.366],0.141[0.248]  & -18.3[-16.5],11.7[13.5],1.41[1.62]         & 7.06[6.15],6.82[5.59],4.02[4.30]  \\
\tableline
\end{tabular}
\label{tab:regions}
\footnotetext{\tiny $^a$All quantities corresponding to the 50-pc region are listed first, with the 25-pc region value in brackets. The central positions of the regions are shown in the 2nd column (chosen by eye for the 50-pc scale; the 25-pc region is centered on the center of mass of the 50-pc region). The 3rd and 4th columns show the total enclosed masses of gas ($M_g$) and stars ($M_*$). Then, for each filament, the upper row shows volume or area-weighted quantities and the lower row shows mass-weighted quantities (indicated by, e.g., $\bar{f_{\Sigma1}})$ . In the 5th and 6th columns are the fractions of the regions with $\Sigma>0.1$ and 1~$\rm g\:cm^{-2}$, respectively, as viewed along the $x,y,z$ axes. The 7th and 8th columns show the mean velocities and velocity dispersions, respectively.}
\footnotetext{\tiny $^b$Only this region has formed stars, with a total mass of 731~$M_\odot$.}
\end{sidewaystable*}




\subsection{Filament Structure}

We study filament properties by defining 10 individual slab regions
(10~pc$\times$10~pc$\times$5~pc) that appear as strips in projection
(10~pc$\times$5~pc) along each filament. These are iteratively
centered on the center of mass within their 10~pc by 10~pc extent in
the plane perpendicular to the main filament axis. Choosing a 10~pc
width for the slabs/strips is a somewhat arbitrary choice. We
therefore also assess inner filament regions of 5~pc width and 5~pc
depth (again iteratively re-centered on the center of mass within
their extent). These outer and inner filament regions are shown in
Figure~\ref{fig:filamentsigma}. We also define a ``dense'' filament as
the material inside the inner filament with $n_{\rm H}\geq 10^3\:{\rm
  cm^{-3}}$.

The mean total filament mass within the 10~pc wide strips is $2.01
\times 10^{5}\:M_{\odot}$: note most of the 50-pc region masses are
contained within these defined filament regions.

We also define ``envelope'' regions that extend half the strip width
on either side of the filament strips (i.e., 5~pc for the outer
filaments and 2.5~pc for the inner filaments). They have the same
depth as the filaments, i.e., 10 and 5~pc for outer and inner cases,
respectively. So with their 5~pc extent along the filament axis, this
gives these total envelope region an equal volume as the filament. We
define the envelope region of the dense filament as the material
inside the inner filament with $n_{\rm H}<10^3\:{\rm cm^{-3}}$.

We calculate physical properties in these filament and envelope
regions, which are listed in Table~\ref{tab:filamentstructure}. These
include volume-averaged densities, which have mean values of $n_{\rm
  H,f} = (3.46,4.26) \times 10^{3}\: {\rm cm}^{-3}$ for the outer and
inner filaments, respectively. Their envelope regions have mean
densities $n_{\rm H,e} = (0.0414,0.449) \times 10^{3}\: {\rm
  cm}^{-3}$, illustrating the decreasing, although still substantial,
density contrast as we zoom into the inner filament regions.

\subsubsection{Longitudinal Structure}

\begin{figure}
\begin{center}$
\begin{array}{cc} 
\includegraphics[width=6.0in]{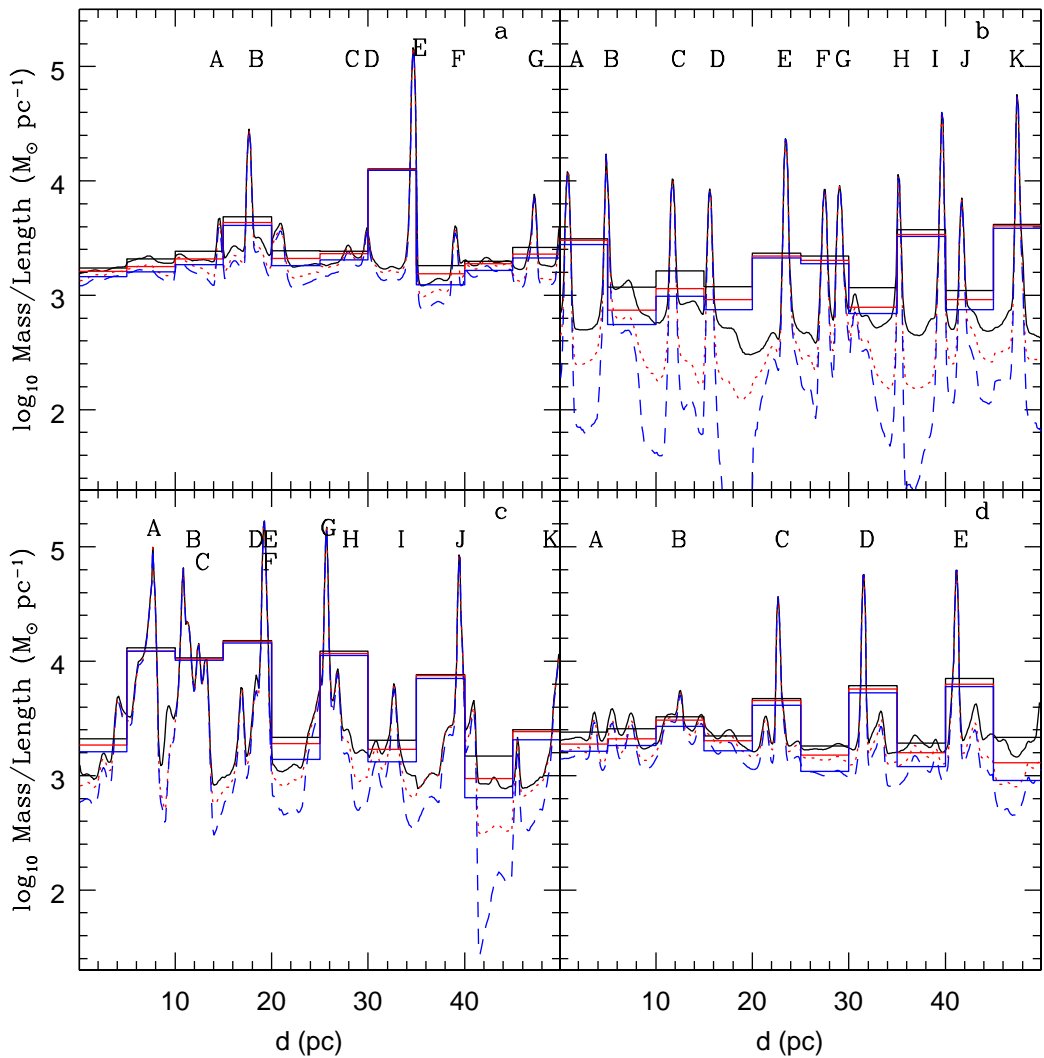}
\end{array}$
\end{center}
\caption{
Mass per unit length profiles along Filaments $a$-$d$, calculated in
$\sim$0.1-pc wide strips perpendicular to the outer filament (black),
inner filament (red), and inner filament where $n_{\rm H} > 10^{3}
cm^{-3}$ (blue).  The mean values for the 10 strips are also plotted
in histogram form. Individual identified clumps (see text) are also
labelled as ``A, B, C, etc.'' Sometimes different clumps can have
similar longitudinal coordinates but are separated laterally, as in
clumps D, E and F in filament $c$.
%
}\label{fig:moverl}
\end{figure}

The mass per unit length, $m_l$, profiles of the filaments are shown
in Figure~\ref{fig:moverl}. Typical median values are $\sim
10^{3.5}\:M_\odot\:{\rm pc}^{-1}$ (about $10^{3}\:M_\odot\:{\rm
  pc}^{-1}$ in filament $b$), but with large fluctuations due to the
formation of dense clumps within the filaments. Thus we also measure
the dispersion in $m_l$, assessed both at the finest resolution
available in the simulation and on the 1~pc scales over the entire
50~pc length of the filaments, and report the values in
Table~\ref{tab:filamentstructure}.

The values can be compared with those measured in IRDCs. Hernandez et
al. (2012) measured $m_l\simeq 300\:M_\odot\:{\rm pc}^{-1}$ along a
3.8~pc length and $\sim 1$~pc wide region of IRDC H from the BT09
sample (G035.30-00.33) (note this region is part of a longer
filamentary structure). Dividing the filament into four strips, these
showed dispersion of about 25\% in their values of $\Sigma$.  The
Orion A filament, studied by Bally et al. (1987), extends over about
13~pc with a similar value of $m_l\simeq 400\:M_\odot\:{\rm pc}^{-1}$.

Longer filaments have been identified and studied by Jackson et
al. (2010), Battersby \& Bally (2012) and Ragan et al. (2014). Jackson
et al. (2010) identified an 80~pc long filament (``Nessie'') with 
typical values of $m_l\sim 500\:M_\odot\:{\rm pc}^{-1}$, estimated
assuming virial equilibrium. Battersby \& Bally (2012) found another
80-pc-long filament, identified in $\rm ^{13}CO$ and with mass $\sim
10^5\:M_\odot$ and $m_l\sim 10^3\:M_\odot\:{\rm pc}^{-1}$. Ragan et
al. (2014) studied seven filaments with typical length $\sim 100$~pc
and average $m_l \simeq 100\:M_\odot\:{\rm pc}^{-1}$.

Comparison of these observed filament properties with those from our
simulation, indicates that the simulated filaments have much larger
values of mass per unit length, by factors of about several to ten. A
systematic and quantitative study of the dispersion in $m_l$ remains
to be carried out for the observed filaments, but initial indications,
e.g., from the Hernandez et al. (2012) study, suggest the observed
filaments have much smaller dispersions in $m_l$ than our simulated
filaments.
\newpage
\clearpage
\begin{sidewaystable*}[counterclockwise]
\setlength\arraycolsep{3pt}

\tiny
\begin{center}
\caption{Filament Structure$^a$
} 
\setlength{\tabcolsep}{0.035in} 
\begin{tabular}{c|c|c|c|c|c|c|c}

\tableline
\tableline
Filament & $\bar{x}_{c}$, $\bar{y}_{c}$, $\bar{z}_{c}$ & $M_g$ & $m_{f}$ & $\sigma_{m_{f}}$ & $n_{\rm H,f}$ & $n_{\rm H,e}$ &  $w_{\rm rms}$
\\
 & (pc) &  (10$^4$ $M_\odot$)   & (10$^4$ $M_{\odot}$/pc) & (10$^4$ $M_{\odot}$/pc) &  ($10^{3}$ cm$^{-3}$) & ($10^{3}$ cm$^{-3}$) & (pc) \\

\tableline
%
\tableline
$a$ &   480, 545, 3.00    & 17.6, 16.3, 15.1    &  0.352, 0.326, 0.303 & 0.337, 0.338, 0.339 & 1.02, 3.78, 18.6      &   0.0149, 0.366, 0.333        &   1.40, 0.801, 0.775           \\
\tableline                                     
$a1$ &  458, 537, 1.48    & 0.868, 0.809, 0.731 & 0.174, 0.162, 0.146 & 0.0264, 0.0246, 0.0226 & 0.502, 1.87, 14.6   & 0.00778, 0.240, 0.204         &  1.67, 0.820, 0.803              \\
$a2$ &  463, 530, 1.81    & 1.04, 0.888, 0.801  & 0.208, 0.178, 0.160 & 0.0324, 0.0286, 0.0278 & 0.603, 2.05, 18.6    & 0.00646, 0.359, 0.223        &  1.60, 0.615, 0.604              \\
$a3$ &  468, 541, 2.54    & 1.21, 1.05, 0.927   & 0.242, 0.209, 0.185 & 0.0760, 0.0693, 0.0678 & 0.699, 2.42, 16.3    & 0.00836, 0.498, 0.316        &  1.51, 0.673, 0.604             \\
$a4$ &  473, 541, 2.55    & 2.42, 2.17, 2.05    & 0.484, 0.435, 0.409 & 0.595, 0.594, 0.593    & 1.40, 5.03, 21.9      & 0.0147, 0.671, 0.373        &  1.24, 0.674, 0.642               \\
$a5$ &  478, 543, 1.73    & 1.22, 1.05, 0.910   & 0.244, 0.210, 0.182 & 0.0908, 0.0870, 0.0860 & 0.707, 2.43, 10.4    & 0.0207, 0.465, 0.407         &  1.58, 1.04, 1.05               \\
$a6$ &  483, 544, 0.939   & 1.21, 1.16, 1.02    & 0.243, 0.232, 0.203 & 0.0616, 0.0604, 0.0592 & 0.705, 2.69, 11.4    & 0.0189, 0.158, 0.423         &  1.35, 0.852, 0.828                \\
$a7$ &  488, 543, 1.38    & 6.41, 6.34, 6.20    & 1.28, 1.26, 1.24    & 3.25, 3.25, 3.24       & 3.71, 14.7 , 57.9      & 0.0339, 0.254, 0.437       &  0.735, 0.514, 0.502                 \\
$a8$ &  493, 541, 1.52    & 0.908, 0.768, 0.622 & 0.182, 0.154, 0.124 & 0.0767, 0.0716, 0.0712 & 0.525, 1.78, 11.6   & 0.0229, 0.403, 0.386          &  1.65, 1.11, 1.08               \\
$a9$ &  498, 541, 0.938   & 0.994, 0.940, 0.823 & 0.199, 0.188, 0.165 & 0.0311, 0.0295, 0.0269 & 0.575, 2.17, 12.3    & 0.00672, 0.110, 0.320        &  1.35, 0.965, 0.932             \\
$a10$ & 503, 539, 1.36    & 1.31, 1.14, 1.06    & 0.263, 0.229, 0.213 & 0.148, 0.147, 0.146    & 0.760, 2.65, 11.3    & 0.00840, 0.501, 0.242        &  1.34, 0.741, 0.711              \\
\tableline                                                           
$b$  &   625, 610, $-$3.50      & 10.9, 9.61, 8.83    &  0.219, 0.192, 0.176   &   0.115, 0.122, 0.121         & 0.633, 2.22, 1.94   &  0.116, 0.382, 0.204   &  1.78, 0.885, 0.786        \\
\tableline                                                                                                                                                   
$b1$ &    625, 588, 1.78      & 1.55, 1.51, 1.40       &  0.311, 0.302, 0.279   &   0.472, 0.469, 0.469       & 0.899, 3.49, 31.4    & 0.109, 0.357, 0.293    &  1.24, 0.578, 0.489           \\
$b2$ &    627, 593, 1.34      & 0.592, 0.371, 0.276    &  0.118, 0.0741, 0.0553 &   0.0962, 0.0971, 0.0966    & 0.342, 0.857, 5.85    & 0.0944, 0.608, 0.245  &  2.41, 1.30, 1.26          \\
$b3$ &    628, 598, 0.868     & 0.820, 0.569, 0.490    &  0.164, 0.114, 0.0979  &   0.237, 0.239, 0.239      & 0.474, 1.31, 12.8    & 0.114, 0.616, 0.200     &  2.31, 0.794, 0.614        \\
$b4$ &    628, 603, 0.124     & 0.594, 0.457, 0.374    &  0.119, 0.0913, 0.0747 &   0.197, 0.193, 0.193     & 0.344, 1.01, 19.3    & 0.123, 0.391, 0.201      &  2.04, 0.790, 0.508         \\
$b5$ &    629, 608, 0.781     & 1.16, 1.11, 1.05       &  0.233, 0.221, 0.211   &   0.531, 0.529, 0.530       & 0.674, 2.56, 22.9    & 0.124, 0.179, 0.139    &  1.20, 0.828, 0.780            \\
$b6$ &    629, 613, 0.717     & 1.10, 1.01, 0.942      &  0.221, 0.202, 0.188   &   0.266, 0.266, 0.267      & 0.639, 2.34, 13.9    & 0.123, 0.253, 0.187     &  1.42, 0.956, 0.902            \\
$b7$ &    628, 618, $-$1.45   & 0.584, 0.395, 0.347    &  0.117, 0.0789, 0.0693 &   0.0467, 0.0344, 0.0349     & 0.337, 0.913, 9.16    & 0.111, 0.509, 0.122  &  2.35, 1.21, 1.17              \\
$b8$ &    628, 623, $-$0.270  & 1.88, 1.70, 1.64       &  0.376, 0.341, 0.327   &   0.830, 0.830, 0.831       & 1.09, 3.94, 32.3    & 0.120, 0.293, 0.179     &  1.53, 0.667, 0.610           \\
$b9$ &    628, 628, $-$0.730  & 0.549, 0.457, 0.373    &  0.110, 0.0914, 0.0746 &   0.151, 0.150, 0.151     & 0.318, 1.06, 7.81    & 0.105, 0.221, 0.219      &  2.05, 0.979, 0.824           \\
$b10$ &   628, 633, $-$0.942  & 2.09, 2.03, 1.94       &  0.418, 0.406, 0.387   &   1.16, 1.15, 1.15        & 1.21, 4.70, 39.0       & 0.137,  0.390, 0.250   &  1.24, 0.741, 0.695            \\
\tableline    
$c$ &   614, 0.251, 20.3  & 34.6, 32.8, 31.1     & 0.692, 0.656, 0.621   & 0.547, 0.537, 0.540     & 2.00, 7.59, 50.6     &  0.0191, 0.491, 0.459          & 1.38, 0.743, 0.645         \\
\tableline                                                                                                                                                        
$c1$ &   613, 229, 14.7  & 1.05, 0.926, 0.803    & 0.210, 0.185, 0.161   & 0.131, 0.123, 0.125  & 6.06, 21.4, 16.5  & 0.0176, 0.364, 0.320      &  1.80, 0.953, 0.755                  \\
$c2$ &   614, 234, 15.7  & 6.53, 6.17, 6.09      & 1.31, 1.23, 1.22      & 1.91, 1.92, 1.92       & 3.77, 14.3, 98.2     & 0.0243, 1.06, 0.208  &  1.02, 0.283, 0.260                  \\
$c3$ &   616, 239, 10.8  & 5.36, 5.23, 5.10      & 1.07, 1.05, 1.02      & 1.30, 1.30, 1.30       & 3.10, 12.1, 109     & 0.0266, 0.338, 0.329  &  0.712, 0.235, 0.199                 \\
$c4$ &   618, 244, 18.9  & 7.59, 7.39, 7.25      & 1.52, 1.48, 1.45      & 3.60, 3.59, 3.59      & 4.39, 17.1, 108     & 0.0299, 0.540, 0.384   &  0.951, 0.575, 0.567                  \\
$c5$ &   613, 249, 21.1  & 1.09, 0.952, 0.692    & 0.217, 0.190, 0.138   & 0.124, 0.126, 0.113 & 0.629, 2.20, 12.9  & 0.0162, 0.348, 0.689      &  2.01, 1.11, 0.913                  \\
$c6$ &   612, 254, 20.8  & 6.13, 5.86, 5.61      & 1.23, 1.17, 1.12      & 2.95, 2.95, 2.94     & 3.54, 13.6, 53.8     & 0.0193, 0.734, 0.750   &  1.03, 0.676, 0.660                 \\
$c7$ &   615, 259, 20.6  & 1.02, 0.848, 0.662    & 0.204, 0.170, 0.132   & 0.131, 0.129, 0.129 & 0.590, 1.96, 20.4  & 0.00959, 0.535, 0.467     &  1.56, 0.798, 0.668                 \\
$c8$ &   615, 264, 24.1  & 3.80, 3.73, 3.52      & 0.761, 0.747, 0.704   & 1.80, 1.79, 1.79     & 2.20, 8.64, 53.9     & 0.00759, 0.245, 0.579  &  0.903, 0.697, 0.678                \\
$c9$ &   614, 269, 24.1  & 0.741, 0.471, 0.320   & 0.148, 0.0943, 0.0640 & 0.110, 0.106, 0.103  & 0.428, 1.09, 15.8  & 0.0328, 0.589, 0.369     &  2.62, 1.25, 0.955                  \\
$c10$ &  613, 274, 27.7  & 1.26, 1.21, 1.02      & 0.252, 0.242, 0.205   & 0.291, 0.291, 0.291   & 0.729, 2.79, 17.3  & 0.00758, 0.157, 0.492   &  1.20, 0.849, 0.800                  \\
\tableline                                                                                                                                                        
$d$ &   785, 200, 10.0   & 17.1, 15.0, 13.2       &  0.342, 0.300, 0.264    & 0.188, 0.185, 0.184      & 0.990, 3.46, 15.9     &  0.0155, 0.557, 0.520    &   1.55, 0.808, 0.772          \\ 
\tableline                                      
$d1$ &  782, 178, 5.98   &  1.20, 0.946, 0.816    &  0.239, 0.189, 0.163    & 0.0530, 0.0492, 0.0484   & 0.692, 2.19, 11.3     &  0.0290, 0.631, 0.363    & 1.78, 0.759, 0.713           \\ 
$d2$ &  783, 183, 6.05   &  1.28, 1.04, 0.919     &  0.257, 0.210, 0.184    & 0.0780, 0.0644, 0.0648   & 0.742, 2.40, 7.80     &  0.0148, 0.630, 0.382    & 1.66, 0.886, 0.854           \\ 
$d3$ &  784, 188, 6.08   &  1.63, 1.52, 1.35      &  0.326, 0.305, 0.270    & 0.0944, 0.0924, 0.0906   & 0.943, 3.52, 10.7     &  0.0198, 0.331, 0.574    & 1.31, 0.814, 0.797            \\ 
$d4$ &  785, 193, 6.12   &  1.11, 1.01, 0.826     &  0.222, 0.202, 0.165    & 0.0445, 0.0417, 0.0376   & 0.642, 2.34, 8.84  &  0.0122, 0.282, 0.548       & 1.42, 0.872, 0.828            \\ 
$d5$ &  786, 198, 6.33   &  2.36, 2.26, 2.06      &  0.471, 0.453, 0.413    & 0.742, 0.742, 0.741      & 1.36, 5.24, 29.0     &  0.0135, 0.311, 0.555     & 1.01, 0.489, 0.455             \\ 
$d6$ &  786, 203, 6.39   &  0.912, 0.753, 0.547   &  0.182, 0.151, 0.109    & 0.0278, 0.0219, 0.0167   & 0.528, 1.74, 12.7   &  0.00820, 0.501, 0.531     & 1.63, 0.788, 0.687            \\ 
$d7$  &  786, 208, 6.84  &  3.06, 2.86, 2.64      &  0.612, 0.572, 0.528    & 1.24, 1.24, 1.24         & 1.77, 6.61, 33.9        &  0.00972, 0.532, 0.616 & 1.17, 0.576, 0.558              \\ 
$d8$  &  785, 213, 7.28  &  0.960, 0.790, 0.599   &  0.192, 0.158, 0.120    & 0.0345, 0.0295, 0.0263   & 0.555, 1.83, 8.38     &  0.0134, 0.424, 0.531    & 1.76, 0.913, 0.877             \\ 
$d9$  &  784, 218, 7.68  &  3.53, 3.16, 2.99      &  0.706, 0.632, 0.598    & 1.34, 1.35, 1.35         & 2.04, 7.31, 32.0        &  0.00491, 0.933, 0.506 & 1.31, 0.665, 0.654              \\ 
$d10$ &  782, 223, 8.97  &  1.08, 0.652, 0.454    &  0.216, 0.130, 0.0907   & 0.0740, 0.0297, 0.0281   & 0.625, 1.51, 4.63     &  0.0292, 0.992, 0.593    & 2.44, 1.32, 1.30              \\ 

\end{tabular}
\end{center}
\footnotetext{\tiny $^a$The 2nd column shows the center of mass of the 10-pc-scale filaments.
From the 3rd column onwards, in each column all quantities are listed
in order of the ``outer'' (10-pc-wide), ``inner'' (5-pc-wide) and
``dense'' (inner filament where $n_{\rm H} > 10^{3}\:{\rm cm^{-3}})$
filaments. The 3rd column lists gas mass, $M_g$.  
The 4th column lists mass per unit length in the filaments, $m_f$,
while the 5th column lists its dispersion, $\sigma_{m_f}$ (note the
values for individual strips are based on the structure seen at the
finest $\sim 0.1$~pc scale, while those for the whole filaments are
based on 50 $\times$ 1-pc-wide regions. The 6th and 7th columns list
mean volume densities in the filament and envelope regions. The 8th
column lists rms lateral widths.
}

\label{tab:filamentstructure}
\end{sidewaystable*}
\clearpage
\subsubsection{Filament Fragmentation}

The peaks in $m_l$ correspond to dense clumps that appear to have
formed by fragmentation of the filament. In gravitationally bound
filaments, clumps are predicted to form by fragmentation at roughly
regular intervals, caused by the so-called sausage-like
fluid instability (e.g., Stodolkiewicz 1964, Nagasawa 1987; Inutsuka \& Miyama 1991)
In our simulated clouds, a large number of dense clumps are observed
to form along the filaments at what appear to be roughly regular
intervals, especially in filaments b and c.

The ability of the simulation to properly resolve fragmentation can be
assessed by reference to the Truelove et al. (1997) criterion, which
requires at least 4 cells per Jeans length, $\lambda_J = (\pi c_{\rm
  th}^2/[G\rho])^{1/2} = 0.92 (c_{\rm th}/0.2\:{\rm km\:s^{-1}})
(n_{\rm H}/10^3\:{\rm cm^{-3}})^{-1/2}\:$pc. Thus with our maxiumum
resolution of 0.122~pc, we are able to resolve fragmentation down to a
level when $\lambda_J=0.49$~pc, i.e., densities of about $n_{\rm
  H}\sim 3500\:{\rm cm}^{-3}$ for sound speeds of 0.2~km~s$^{-1}$. 
This is a relatively low density compared to the “clump” material defined in Paper I at a density threshold of $10^5\:{\rm cm^{-3}}$.
 Note that the TT09 simulation imposed an effective temperature floor of 300 K, corresponding to an effective sound speed of 1.8 km~s$^{-1}$, 
to mimic effects of microturbulence. The actual fragmentation of the GMCs and filaments involves gas that is quickly cooling from a few 
hundred K down towards ∼ 10 K, but with the structures also buffeted by turbulent motions (with speeds from 1 to 10 km~s$^{-1}$) imparted from the
 cloud bulk motions and their gravitational collapse. Given such conditions, we cannot be confident that the actual fragmentation we see in 
the simulation is free of numerical artifacts, and so the following properties of the spacing of the clumps should be treated with caution. 
Comparing the 0.5-pc resolution simulations
with the 0.1-pc resolution ones, we do identify most of the same
clumps forming in the same locations, with the fragmentation of only a
few of smaller clumps unresolved. However, given the above
considerations, we cannot be fully confident that the actual
fragmentation we see in the simulation is free of numerical artifacts,
and so the following properties of the spacing of the clumps should be
treated with caution.

To investigate clump spacing quantitatively, we identify clumps using
the clump-finding routine of Smith et al. (2008) that is within the
{\it yt} software package (Turk et al. 2011). This routine identifies
topologically connected structures using a recursive contouring
algorithm, given a density range and density increment.  One can also
specify a minimum number of contiguous cells to be considered a clump.
We set this threshold to 50 cells. We choose a density increment of
one order of magnitude in $n_{\rm H}$, ranging from $n_{\rm H} =
10^{5}\: {\rm cm}^{-3}$ to $n_{\rm H} = 10^{8}\: {\rm cm}^{-3}$, near
the maximum density reached by our simulation. The routine identifies
7, 11, 11 and 5 clumps in filaments $a$--$d$, respectively. The 3D
distance between each clump center and the nearest clump along the
filament is calculated.  The mean separations in each filament are
5.30, 5.26, 5.87, and 7.84~pc with dispersions of 3.42, 3.27, 4.29,
and 4.56~pc for filaments $a$--$d$, respectively.  We find a mean
separation for all clumps of 5.54~pc, with dispersion of 3.60~pc.

Note that the number and properties of the clumps are sensitive to the
choice of threshold density and the required minimum number of
cells. Considering filament $c$, if the density threshold is
raised and lowered by a factor of two, then the number of clumps changes
from 11 to 9 and 15, respectively. If the minimum required
number of cells is raised/lowered by a factor of two, then the number
of clumps changes from 11 to 9 and 15, respectively.

Fragmentation at regular intervals has been seen in large filamentary
IRDCs. For example, Jackson et al. (2010) find a fairly regular
spacing of $\sim 4.5\:{\rm pc}$ between individual dense clumps. While
this qualitative agreement is noteworthy, given the numerical
difficulties of properly resolving fragmentation and the sensitivity
of results to clump identification method parameters, we cannot draw
any firm conclusion from this result. Still, the fragmentation spacing
we observe in the simulation serves as a point of comparison for
future numerical studies that achieve higher resolution and that
include magnetic fields.

\clearpage
\begin{sidewaystable*}[counterclockwise]
\setlength\arraycolsep{2pt}

\tiny
\caption{Filament Kinematics} 
\setlength{\tabcolsep}{0.01in} 
\begin{tabular}{c|c|c|c|c|c|c|c|c|c|c|c}

\tableline
\tableline
Filament &   $\bar{v}_{\rm x}$ $^{a}$  & $\bar{v}_{\rm y}$ $^{a}$ & $\bar{v}_{\rm z}$ $^{a}$ &  $\bar{\sigma}_{f,x}$ &   $\bar{\sigma}_{f,y}$ &   $\bar{\sigma}_{f,z}$ & $\bar{\sigma}_{e,x}$  & $\bar{\sigma}_{e,y}$  & $\bar{\sigma}_{e,z}$  & $\rm m_{\rm f}/\rm \bar{m}_{\rm vir}$ & $log_{10} P_{e}/P_{f}$  \\
  &  (km\ s$^{-1}$)  & (km\ s$^{-1}$) & (km\ s$^{-1}$) & (km\ s$^{-1}$) & (km\ s$^{-1}$) & (km\ s$^{-1}$) &  (km\ s$^{-1}$) &  (km\ s$^{-1}$) &  (km\ s$^{-1}$)  &  & \\

\tableline
$a$   & -1.33,-1.51,-0.974   & -0.223,0.242,0.531 & 0.424,0.430,0.444 & 4.16,4.09,3.47 & 4.75,4.61,4.24 & 2.79,2.78,2.68 & 2.20,2.49,2.79 & 5.23,2.83,5.02 & 2.51,1.54,2.43 & 0.269,0.903,0.236 &-1.66,-1.36,-1.43   \\
\tableline

$a1$  & -8.03, -8.08, -7.77  & 2.85, 2.92, 3.11                & 0.283, 0.268, 0.208            & 3.37, 3.19, 2.95   & 2.42, 2.46, 2.30 & 0.855, 0.652, 0.502  & 2.11, 1.21, 2.16 & 4.16, 1.50, 3.73 & 2.28, 1.25, 1.30 & 0.638, 0.574, 0.596  & -1.34, -1.32, -1.58   \\
$a2$  & -6.65, -7.39, -6.96  & 3.55, 3.97, 4.22                & 0.186, 0.222, 0.158            & 4.16, 4.05, 3.83   & 2.56, 2.47, 2.33 & 0.957, 0.814, 0.714  & 2.80, 1.27, 2.41 & 5.53, 1.49, 3.83 & 2.92, 1.19, 1.29 & 0.682, 0.628, 0.634  & -1.30, -1.19, -1.86   \\
$a3$  & -3.15, -3.33, -2.74  & 2.52, 2.95, 3.48                & -0.0215, -0.0810, -0.122       & 4.22, 4.26, 3.81   & 4.13, 4.22, 3.92 & 1.63, 1.63, 1.63   & 2.74, 1.10, 1.60 & 5.08, 2.22, 4.75 & 2.83, 1.02, 1.59   & 0.304, 0.253, 0.259  & -1.74, -1.24, -1.67   \\
$a4$  & -3.88, -4.25, -4.05  & 3.91, 4.62, 4.96                & -0.526, -0.584, -0.673         & 4.18, 4.15, 4.01   & 5.53, 5.31, 5.15 & 4.25, 4.46, 4.53   & 2.82, 2.31, 2.05 & 4.61, 3.12, 4.96 & 3.02, 1.24, 2.70   & 0.341, 0.331, 0.332   & -2.13, -1.34, -1.81   \\
$a5$  & -3.30, -3.77, -3.18  & 0.471, 1.53, 1.98               & 0.485, 0.587, 0.557            & 4.64, 4.73, 4.38   & 4.65, 3.84, 3.49 & 2.15, 2.25, 2.21   & 2.60, 1.97, 2.08 & 6.02, 2.96, 5.12 & 2.93, 1.29, 2.47   & 0.243, 0.307, 0.321  & -1.31, -0.942, -1.16   \\
$a6$  & 1.93, 2.21, 3.17     & -1.45, -1.04, -0.371            & 0.433, 0.349, 0.373            & 4.05, 3.85, 2.25   & 4.80, 4.47, 3.65 & 1.95, 1.86, 1.74   & 2.49, 2.28, 2.28 & 4.98, 3.71, 5.62 & 2.37, 1.82, 2.51   & 0.227, 0.250, 0.329  & -1.53, -1.39, -0.939  \\
$a7$  & 1.48, 1.53, 1.66     & 1.94, 2.04, 2.26                & -0.312, -0.335, -0.348         & 5.36, 5.35, 5.28   & 10.2, 10.2, 10.1 & 9.79, 9.84, 9.93   & 2.21, 3.21, 2.93 & 6.08, 3.98, 5.37 & 2.92, 2.31, 3.63   & 0.263,  0.261, 0.258  & -2.49, -2.58, -2.31   \\
$a8$  & -2.15, -2.82, -2.04  & -4.57, -3.74, -3.75             & 1.62, 1.74, 2.13               & 4.56, 4.39, 3.92   & 5.27, 5.20, 4.67 & 2.48, 2.51, 2.15   & 1.77, 4.18, 4.21 & 5.59, 3.01, 4.77 & 1.94, 1.69, 3.19   & 0.141, 0.122, 0.122 & -1.31, -1.18, -1.12   \\
$a9$  & 2.91, 2.99, 3.75     & -5.90, -5.76, -5.73             & 0.923, 0.884, 0.920            & 3.54, 3.41, 2.21   & 3.55, 3.52, 2.81 & 1.65, 1.50, 1.22   & 1.31, 4.96, 4.34 & 5.77, 3.45, 5.08 & 2.04, 2.24, 2.74   & 0.339, 0.325, 0.447  & -1.51, -1.32, -0.838   \\
$a10$ & 7.51, 7.78, 8.42     & -5.55, -5.07, -4.85             & 1.17, 1.25, 1.24               & 3.56, 3.55, 2.06   & 4.38, 4.45, 3.96 & 2.19, 2.25, 2.19   & 1.19, 2.40, 3.88 & 4.49, 2.87, 6.97 & 1.88, 1.38, 2.90   & 0.295, 0.249, 0.291  & -1.93, -1.10, -1.06   \\
  
\tableline
$b$ &    -0.0154,-0.237,-0.233 & -0.638,-0.776,-0.858 & 0.420,0.228,0.170 & 3.37,3.31,3.31 & 1.87,1.94,1.91 & 3.25,2.46,3.52 & 3.41,2.13,2.88 & 1.10,1.09,1.83 & 1.15,1.28,2.64 & 0.563,1.12,0.490 & -0.666,-1.11,-2.03 \\
\tableline
$b1$ &   -2.43, -2.31, -2.29     & -6.64, -6.68, -6.74      & 0.239, 0.271, 0.237         &  3.56, 3.57, 3.64    & 2.06, 2.08, 2.09 & 3.48, 3.60, 3.63  &     3.97, 1.57, 2.55  &  1.69, 1.45, 1.78  & 1.10, 1.20, 3.18  &    0.526, 0.509, 0.453     & -0.822, -1.70, -2.34       \\
$b2$ &   -2.20, -2.36, -2.43     & -6.40, -7.41, -7.97      & 0.802, 0.136, 0.0376       &  2.82, 2.46, 2.63    & 2.65, 2.84, 2.88 & 3.12, 3.65, 3.85  &     3.53, 1.16, 1.84  &  1.18, 0.936, 1.98  & 1.26, 1.18, 2.98  &    0.321, 0.264, 0.172    & -0.364, -0.800, -1.69        \\
$b3$ &   -0.636, -1.15, -1.08    & -3.86, -4.06, -4.11      & 0.717, 0.268, 0.236         &  2.69, 2.46, 2.56    & 1.38, 1.57, 1.61 & 2.56, 2.87, 2.93  &     3.20, 1.08, 1.70  &  0.941, 0.656, 1.24  & 0.908, 1.17, 2.52  & 0.485, 0.401, 0.321     & -0.471, -1.05, -2.16        \\
$b4$ &   -0.390, -0.745, -0.660  & -3.23, -3.42, -3.69      & 0.273, -0.0650, -0.263      &  2.42, 2.47, 2.39    & 1.46, 1.56, 1.45 & 2.23, 2.40, 2.40  &     4.30, 1.96, 2.77  &  1.21, 0.972, 1.52  & 1.24, 1.04, 2.15  &   0.436, 0.321, 0.281     & 0.0514, -0.632, -1.86        \\
$b5$ &   -0.153, 0.102, -0.130   & 0.817, 0.871, 0.897      & 0.211, 0.167, 0.114         &  3.89, 3.98, 3.98    & 1.86, 1.88, 1.87 & 3.90, 3.99, 4.04  &     3.81, 2.80, 3.87  &  1.07, 1.01, 1.94  & 1.30, 1.14, 2.59  &    0.331, 0.301,  0.286     & -0.751, -1.46, -2.24        \\
$b6$ &   0.612, 0.596, 0.624     & -0.0180, -0.0627, -0.0674 & 0.330, 0.280, 0.246        &  2.52, 2.47, 2.43    & 1.70, 1.76, 1.78 & 2.47, 2.55, 2.53  &     3.72, 1.97, 3.02  &  0.890, 0.727, 1.48  & 1.23, 1.10, 2.77  &  0.745, 0.710, 0.688     & -0.376, -1.16, -1.68        \\
$b7$ &   1.69, 1.33, 1.42        & 2.91, 3.14, 3.37         & 0.803, 0.768, 0.718         &  2.81, 2.82, 2.62    & 1.73, 1.88, 1.80 & 2.07, 2.35, 2.42  &     4.33, 2.20, 3.88  &  0.996, 1.38, 1.56  & 0.958, 1.45, 1.79  &  0.317, 0.214, 0.217   & -0.108, -0.466, -1.54        \\
$b8$ &   0.773, 0.396, 0.411    & 4.16, 4.32, 4.35         & -0.117, -0.237, -0.275      &  4.59, 4.66, 4.71    & 1.85, 1.85, 1.85 & 4.23, 4.41, 4.45  &     2.67, 2.61, 3.17  &  0.959, 1.04, 1.91 & 1.26, 1.44, 3.07  &     0.383, 0.338, 0.317     & -1.43, -1.63, -2.60       \\
$b9$ &   1.46, 0.686, 0.731       & 1.65, 1.30, 1.08         & 0.777, 0.554, 0.529         &  2.81, 2.61, 2.50    & 1.70, 1.60, 1.53 & 2.50, 2.68, 2.73  &     2.49, 3.61, 3.06  &  0.971, 0.974, 1.58 & 1.00, 1.40, 2.45  &  0.300, 0.288, 0.257   & -0.584, -0.400, -1.37        \\
$b10$ &  1.12, 1.09,  1.07       & 4.23, 4.24, 4.30         & 0.169, 0.134, 0.123         &  5.55, 5.58, 5.68    & 2.32, 2.34, 2.26 & 5.97, 6.06, 6.17  &     2.03, 2.38, 2.93  &  1.08, 1.76, 3.29  & 1.22, 1.69, 2.86  &    0.292, 0.280, 0.258     & -1.81, -1.82, -2.77        \\
 
\tableline
$c$  & -1.16,-0.584,-2.93 & -0.144,-0.0736,-1.60 & -0.635,-0.718,0.310 & 7.14,6.99,6.95 & 5.28,5.31,5.21 & 4.96,5.04,5.02 & 2.91,3.39,4.75 & 2.27,2.82,4.76 & 1.82,1.74,3.45 & 1.83,1.28,0.570 &-2.67,-1.67,-2.23 \\
\tableline 
$c1$ &  4.56, 5.41, 2.90    & -1.86, -1.33, 1.25   & 1.58, 1.61, 1.62           & 5.54, 5.24, 4.96   & 4.31, 4.24, 4.32  & 1.92, 1.97, 1.84 & 2.84, 3.13, 4.09 & 2.74, 2.24, 3.71 & 1.44, 1.43, 2.67  & 0.147, 0.145, 0.140 & -2.12, -1.22, -1.88 \\
$c2$ &  1.24, 1.59, -0.952  & 2.59, 3.12, -0.708   & 0.956, 0.943, 3.20         & 9.77, 9.90, 9.93   & 10.7, 10.7, 10.8  & 3.15, 3.21, 3.17 & 3.91, 3.98, 4.29 & 3.19, 2.87, 6.10 & 2.58, 1.84, 4.21  & 0.294, 0.271, 0.265   & -2.99, -1.92, -3.40 \\
$c3$ &  3.19, 3.37, 1.31    & -0.288, -1.49, -3.48 & -0.0597, -0.131, 0.430     & 8.35, 8.36, 8.38   & 8.57, 8.62, 8.66  & 2.85, 2.84, 2.80 & 3.18, 2.68, 5.06 & 3.16, 2.93, 5.07 & 1.73, 1.31, 3.88  & 0.331, 0.322, 0.312   & -2.90, -2.54, -2.96 \\
$c4$ &  1.47, 1.57, -1.25   & 0.124, 0.147, -1.80  & -0.0681, -0.135, 0.143     & 10.1, 10.2, 10.3   & 3.91, 3.81, 3.75  & 9.30, 9.41, 9.48 & 2.93, 4.94, 6.45 & 2.73, 4.38, 5.91 & 1.41, 1.37, 3.86  & 0.316, 0.303, 0.294   & -3.24, -2.13, -2.85 \\
$c5$ &  -4.19, -4.22, -4.99 & -3.31, -3.23, -4.04  & -1.28, -1.84, -0.722       & 5.01, 4.91, 4.57   & 4.28, 4.32, 4.28  & 4.10, 3.98, 4.06 & 3.80, 4.31, 5.06 & 2.76, 3.13, 4.44 & 2.49, 1.95, 3.73  & 0.187, 0.170, 0.143 & -1.83, -0.915, -1.18 \\
$c6$ &  1.27, 1.52, -1.45   & -2.40, -2.31, -4.64  & -0.708, -0.765, 1.10       & 8.85, 8.92, 9.03   & 4.42, 4.43, 4.43  & 8.42, 8.60, 8.75 & 2.33, 3.64, 5.20 & 1.28, 2.98, 3.87 & 1.50, 1.87, 3.78  & 0.336, 0.317, 0.296   & -3.42, -2.05, -2.34 \\
$c7$ &  -3.95, -3.04, -4.97 & -2.45, -2.55, -4.11  & 0.506, 0.616, 1.50         & 4.01, 3.61, 3.69   & 2.72, 2.80, 2.57  & 2.06, 2.18, 2.15 & 2.32, 3.16, 3.23 & 1.85, 2.94, 3.33 & 1.62, 1.25, 2.24  & 0.272, 0.280, 0.209 & -2.26, -0.680, -1.78 \\
$c8$ &  -4.19, -4.20, -7.76 & 3.45, 3.42, 0.149    & 0.160, 0.223, -0.249       & 7.21, 7.25, 7.37   & 3.00, 3.01, 2.87  & 6.64, 6.68, 6.84 & 2.51, 2.36, 3.93 & 2.37, 1.82, 4.33 & 2.06, 1.70, 2.98  & 0.314, 0.305, 0.279   & -3.38, -2.52, -2.52 \\
$c9$ &  -5.47, -2.20, -3.99 & 2.29, 2.77, 0.533    & 0.968, 1.07, 2.45          & 6.12, 5.02, 4.81   & 3.59, 4.04, 3.99  & 2.50, 2.87, 3.01 & 2.59, 3.00, 4.70 & 1.08, 2.03, 3.98 & 1.38, 2.05, 2.54  & 0.0850, 0.0805, 0.0595 & -1.86, -0.715, -1.65 \\
$c10$ & -5.50, -5.64, -8.13 & 0.410, 0.717, 0.853  & -8.40, -8.77, -6.37        & 6.39, 6.51, 6.49   & 7.31, 7.12, 6.47  & 8.68, 8.62, 8.07 & 2.72, 2.66, 5.52 & 1.50, 2.88, 6.83 & 1.94, 2.60, 4.57  & 0.133, 0.123, 0.105  & -2.73, -2.03, -1.69 \\
                        
\tableline   
$d$  & 1.48,1.39,2.48 & 0.875,0.810,0.156 & 0.239,-0.321,0.419 & 5.63,5.49,3.73 & 5.01,4.96,3.26 & 3.11,3.23,3.24 & 2.49,4.25,6.81 & 3.21,3.93,6.59 & 1.28,5.01,1.78 & 1.67,0.392,0.133 & -2.61,-0.994,-0.841 \\
\tableline                                                       
$d1$ &  3.83, 3.97, 6.21      &  0.609, 0.986, 0.614       &   0.732, 1.01, 1.36           &  5.61, 5.40, 3.29     &  3.56, 3.40, 1.85     &   1.76, 1.78, 1.71   &   1.93, 4.15, 7.63     &   3.02, 2.73, 5.01    &  1.21, 1.03, 1.96     &    0.163, 0.140, 0.324    & -2.30, -0.768, -0.761                   \\
$d2$ &  3.97, 4.10, 5.46      &  0.625, 0.901, 1.30        &   0.546, 0.761, 0.638         &  5.04, 4.83, 2.48      &  3.41, 3.38, 1.79     &   1.89, 1.97, 1.90   &   1.55, 4.08, 8.32     &   2.79, 2.85, 6.10    &  1.18, 4.83, 1.87     &    0.217, 0.192, 0.643    & -2.73, -0.727, -0.258                    \\
$d3$ &  3.52, 3.73, 4.65      &  1.58, 1.83,  1.76         &   0.365, 0.379, 0.409         &  4.59, 4.30, 2.20      &  3.93, 3.71, 2.20     &   2.22, 2.27, 2.26   &   1.74, 4.39, 6.84     &   2.81, 3.20, 5.52    &  1.17, 4.30, 1.91     &    0.333, 0.354, 1.19     & -2.52, -1.01, -0.286                  \\
$d4$ &  2.58, 2.97, 4.23      &  0.769, 1.15,  1.99        &   0.298, 0.318, 0.243         &  5.43, 4.96, 2.24      &  4.53, 4.13, 2.25     &   1.87, 1.89, 1.69   &   1.71, 4.75, 6.47     &   2.74, 3.80, 6.14    &  1.23, 4.96, 1.50     &    0.162, 0.177, 0.705     & -2.72, -0.956, -0.288                    \\
$d5$ &  2.85, 2.96, 3.53      &  2.92, 3.13, 2.00          &   0.352, 0.351, -0.0513       &  5.67, 5.55, 4.52      &  5.13, 4.94, 3.74     &   4.28, 4.36, 4.48   &   1.60, 4.62, 6.20     &   2.57, 3.96, 6.09    &  1.10, 5.55, 1.73     &    0.315, 0.316, 0.434     & -3.10, -1.39, -1.44                    \\
$d6$ &  0.106, -0.191, 2.41   &  0.0161, 0.0646, 1.85      &   0.0292, 0.0430, 0.249       &  6.23, 6.18, 2.86      &  6.37, 6.47, 3.30     &   1.88, 1.95, 1.68   &   1.81, 4.49, 6.09     &   2.18, 4.69, 7.02    &  1.07, 6.18, 1.64     &    0.101, 0.0848, 0.288   & -2.88, -0.820, -0.721                    \\
$d7$  & 1.42, 1.49, 2.35      &  2.45, 2.60, -0.561        &   0.0880, 0.121, 0.181        &  6.63, 6.71, 5.98      &  5.77, 5.75, 4.31     &   5.80, 5.98, 6.18   &   4.61, 3.98, 6.58     &   3.17, 4.62, 6.54    &  1.41, 6.71, 1.76     &    0.299, 0.273, 0.317     & -2.58, -1.55, -1.66                    \\
$d8$  & -2.40, -2.79, -0.260  &  -1.13, -1.61, 0.513       &   -0.0670,-0.00390, 0.474     &  5.77, 5.85, 4.22      &  7.05, 7.33, 4.88     &   2.09, 2.17, 1.99   &   5.77, 4.21, 6.31     &   4.46, 3.77, 8.37    &  1.90, 5.85, 1.81     &    0.124, 0.0993 0.144   & -1.62, -0.919, -0.849                    \\
$d9$  & -0.306, -0.235, -1.22 &  1.39, 1.44, -3.81         &   0.0831, 0.140, 0.522        &  6.55, 6.76, 6.45      &  5.90, 5.98, 5.07     &   6.10, 6.42, 6.58   &   3.42, 3.86, 7.02     &   4.74, 4.87, 7.60    &  1.42, 6.76, 1.87     &    0.353, 0.297, 0.309     & -3.18, -1.38, -1.73                    \\
$d10$ & -4.65, -6.30, -5.34   &  -2.48, -4.81, -3.74       &   -0.345, -6.01, 0.217        &  4.94, 4.74, 3.56      &  6.53, 6.89, 4.83     &   2.19, 2.45, 2.66   &   4.07, 3.94, 6.11     &   4.86, 4.66, 9.33    &  1.69, 4.74, 1.81     &    0.190, 0.125, 0.154  & -1.50, -0.343, -0.422                    \\
%

\end{tabular}
\label{tab:filamentkinematics}
$ $
$ $
$^a$Mass-weighted mean velocities
\end{sidewaystable*}
\clearpage

\subsubsection{Lateral Structure}

\begin{figure}
\begin{center}$
\begin{array}{cc} 
\includegraphics[width=6.0in]{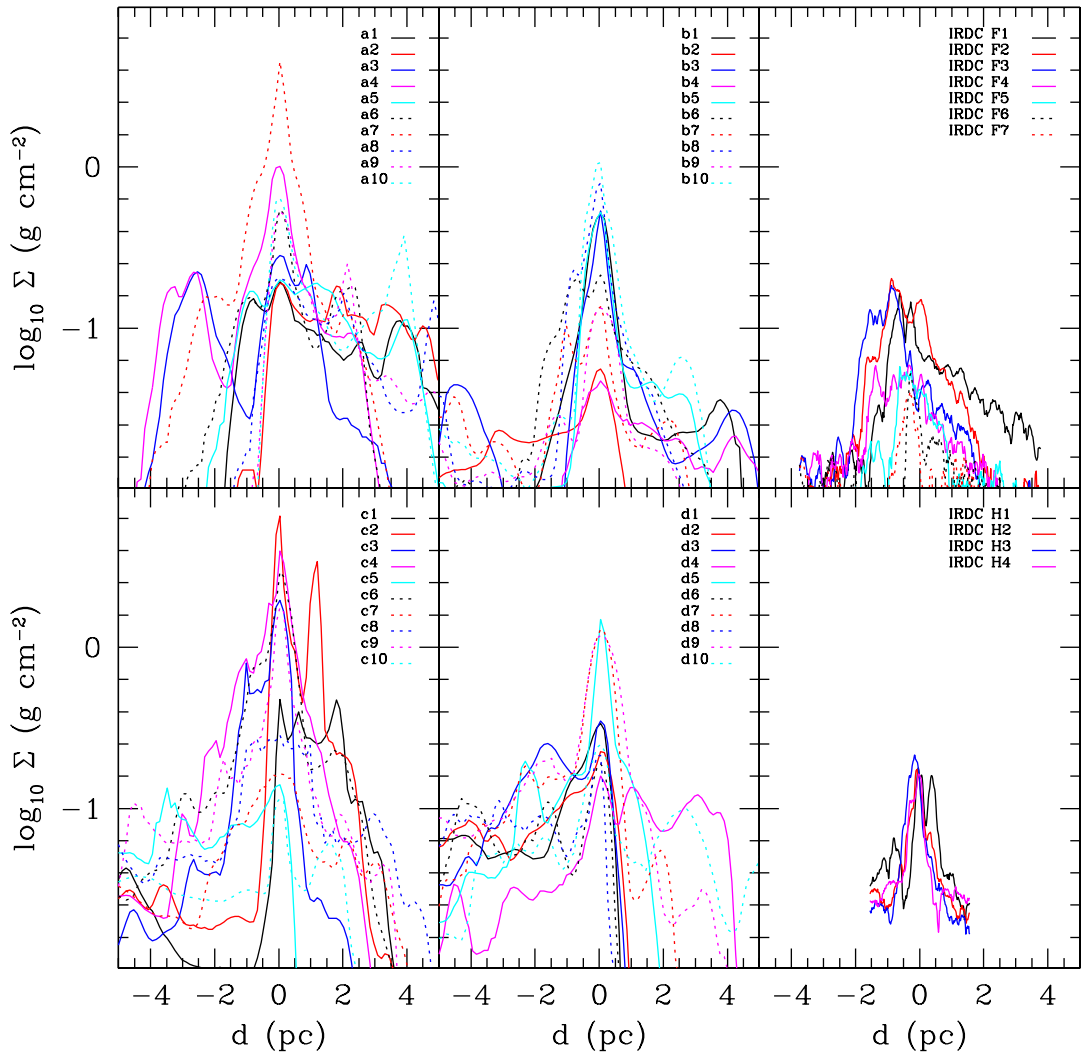}
\end{array}$
\end{center}
\caption{
Mean lateral $\Sigma$ profiles perpendicular to the axes of Filaments
$a$--$d$, calculated in each of the 10 outer filament strip regions,
i.e., 10~pc wide (perpendicular to the filament long-axis), 10~pc
deep, and averaging over a 5~pc length along the filament long-axis.
In the right column we show the profiles across two filamentary IRDCs,
using the strip locations and sizes from Hernandez et
al. (2011).}\label{fig:filwidth}
\end{figure}

The width of filaments in the ISM can provide insight into the
conditions from which they formed.  Arzoumanian et al. (2011) found a
characteristic width of $\sim 0.1$~pc for a sample of 27 filaments in
the {\it Herschel} Gould Belt Survey toward the IC 5146 molecular
cloud, suggesting that the dissipation of large-scale turbulence to
leave thermally-dominated structures may have played an important role
in the filaments' formation (however, see Smith et al. 2014,
whose simulations show that column density filaments are not always
part of a single coherent 3D structure). It remains to be established if such a
common scale of filament width is relevant to more massive IRDCs,
although inspection of the BT12 maps of 10 IRDCs, suggests there is a
range of widths, extending to larger values.

Our simulated filaments, with minimum resolution of about 0.1~pc, are
not well-resolved enough to measure scales of filament width down to
the level claimed by Arzoumanian et al. (2011). For each of the
10~pc-wide and deep strips covering our simulated filaments, the mean
$\Sigma$ profiles are calculated perpendicular to the filament axis
and displayed in Figure~\ref{fig:filwidth}. As can be seen from this
figure and also from Figure~\ref{fig:filamentsigma}, a variety of
profiles are present. The peak $\Sigma$ values range from $\sim 0.1$
to greater than 1~$\rm g\:cm^{-2}$, averaged over the 5-pc-wide extent
of the strip regions. Multiple peaks can be present, sometimes due to
multiple diffuse sub-filaments (e.g., $a1$, $a2$, $a3$) or multiple clumps
(e.g., $a4$). The peaks are often dominated by the presence of a single
clump and the overall profile can be affected by how clumps happen to
be distributed in these strip regions (e.g., $c2$).

In Figure~\ref{fig:filwidth} we also show the lateral $\Sigma$
profiles of strips from IRDC filaments F and H that were defined by
Hernandez \& Tan (2011), but using the latest combined MIR+NIR
extinction maps of Kainulainen \& Tan (2013). One notices that the
peak $\Sigma$ values are smaller in the observed filaments, but a
range of widths are present that is similar to some of the peaks shown by
the simulated filaments. The four strips of IRDC H show relatively
similar lateral profiles, while IRDC F shows a wider variety, more
similar to the simulated filaments.

For a more quantitative comparison we evaluate the mass-weighted ``rms
lateral width,'' $w_{\rm rms}$, i.e., the rms displacement of filament
strip material from its center of mass in the lateral direction. On
scales of the 5~pc-wide strips extending 10~pc laterally, the
simulated filaments have average ($\pm$ dispersion) rms lateral widths
of $1.40\pm0.264$~pc, $1.78\pm0.475$~pc, $1.38\pm0.572$~pc and
$1.55\pm0.385$~pc for $a$--$d$, respectively. On scales of 1~pc-wide
strips the rms lateral widths are $1.06\pm0.302$~pc,
$1.46\pm0.718$~pc, $0.951\pm0.575$~pc and $1.39\pm0.451$~pc. Note this
is the rms width of material within $\pm 5$~pc of the center of mass
of each strip.

We also calculate this width for the IRDCs F and H using the strips
defined in Hernandez et al. (2011) strips, which have widths of $\sim 7.5$~pc and $\sim 3.1$~pc,respectively.  For IRDC F, we find rms
lateral widths of $0.719\pm0.517$~pc, and for IRDC H we find
$0.409\pm0.317$~pc. We also consider the $\pm 5$~pc scale, finding similar widths of $0.657\pm0.482$~pc and $0.406\pm0.290$~pc for IRDCs F and H, respectively.  These are factors of a few smaller than the
simulated filament at the 1~pc-wide strip scale.  These results,
including for indivdiual strips and for inner and dense filaments, are
also listed in Table~\ref{tab:filamentstructure}.

\subsection{Filament and Clump Kinematics}\label{S:kinematics}


\begin{figure}
\begin{center}$
\begin{array}{c} 
\includegraphics[width=7.0in]{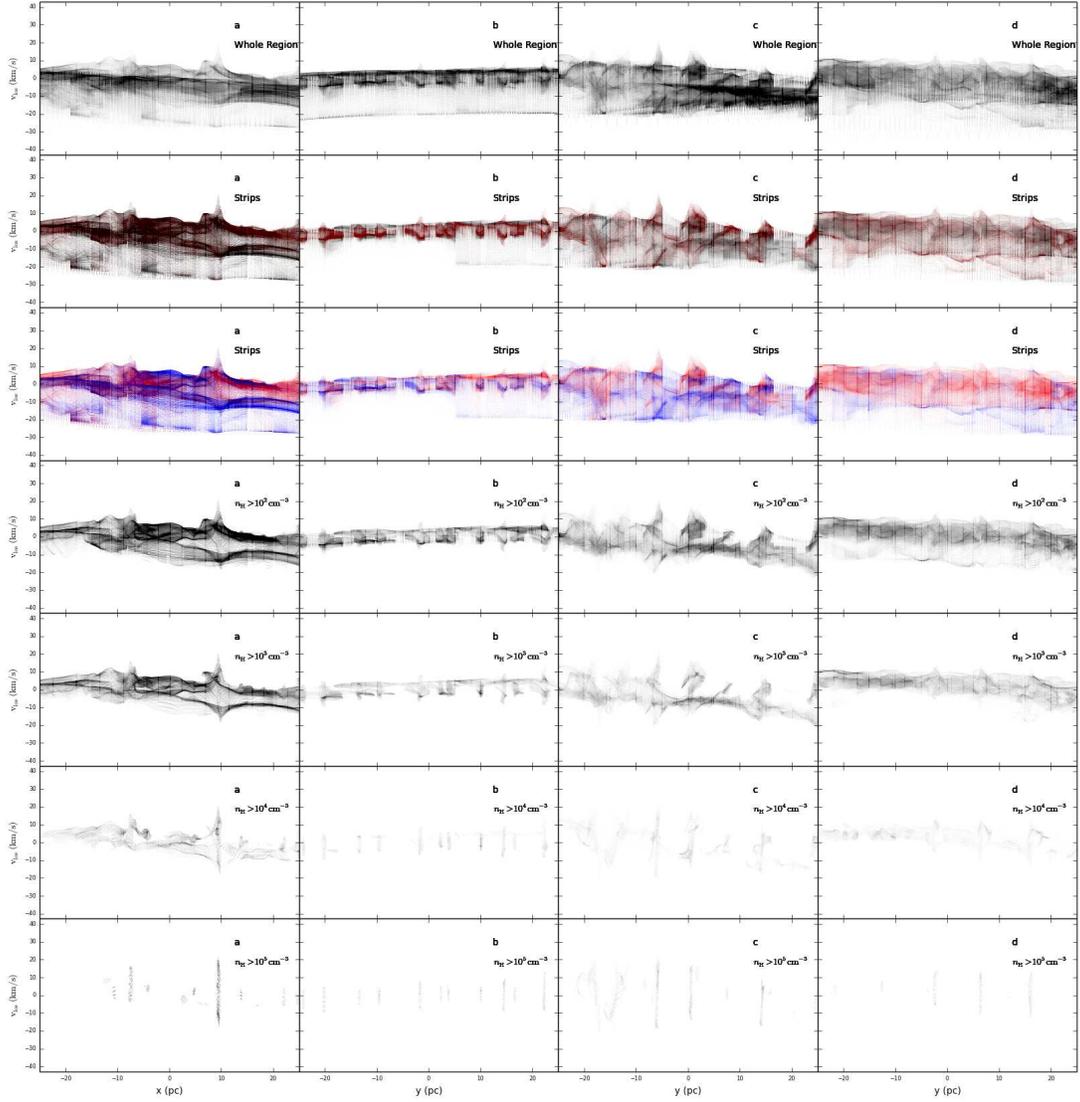}
\end{array}$
\end{center}
\caption{
Position-velocity diagrams for Filaments $a$--$d$ (left to right),
where the position coordinate has been chosen to be along the main
axis of the filament, i.e., $x$-direction for $a$ and $y$-direction
for $b$--$d$. The top row shows all the gas in the 50~$\rm pc^{3}$
regions. For clarity of highlighting kinematic structures, each
simulation cell is marked by a simple dot (thus the AMR gridding
appears as vertical stripes). The second row shows only the material
in the filament strips, with outer (10-pc-wide) strips in black and
inner (5-pc-wide) strips in red. The third row shows material on
near side of the outer filament in red and the far side in blue. The
fourth, fifth, sixth and seventh rows show only gas at $n_{\rm H} >
10^{2}, 10^{3}, 10^{4}$, and $10^{5}\:{\rm cm^{-3}}$, respectively.
%
}\label{fig:posvel}
\end{figure}

We show the position-velocity ($p$-$v$) diagrams for Filaments $a$--$d$ in
Figure~\ref{fig:posvel}, with the position coordinate ranging along
the 50-pc length of the main filament axis and the velocity being the
line of sight velocity if viewing this filament in the Galactic plane.
The figure first shows these diagrams for all the gas in the 50-pc$^3$
regions. A variety of gas distributions are seen (note that to
highlight individual kinematic features, independent of density, each
simulation cell is plotted with an equally-weighted dot). Some
regions, e.g., $a$, $c$ and $d$, show a very broad distribution of
velocities, extending over ranges of $\sim$30~km~s$^{-1}$ or more. Filament
$b$ has a narrower range. Within each region, more coherent structures
can be seen in $p$-$v$ space, including those that
correspond to the identified filaments, which are distinguished
in the second row of the figure (outer filament in black, inner in
red). However, again, there is a variety of velocity structures
exhibited by the different filaments, with $a$, $c$ and $d$ again being more
disordered. This is consistent with the velocity fields shown in
Figure~\ref{fig:nHv}.

\begin{figure}
\begin{center}$
\begin{array}{c} 
\includegraphics[width=6.0in]{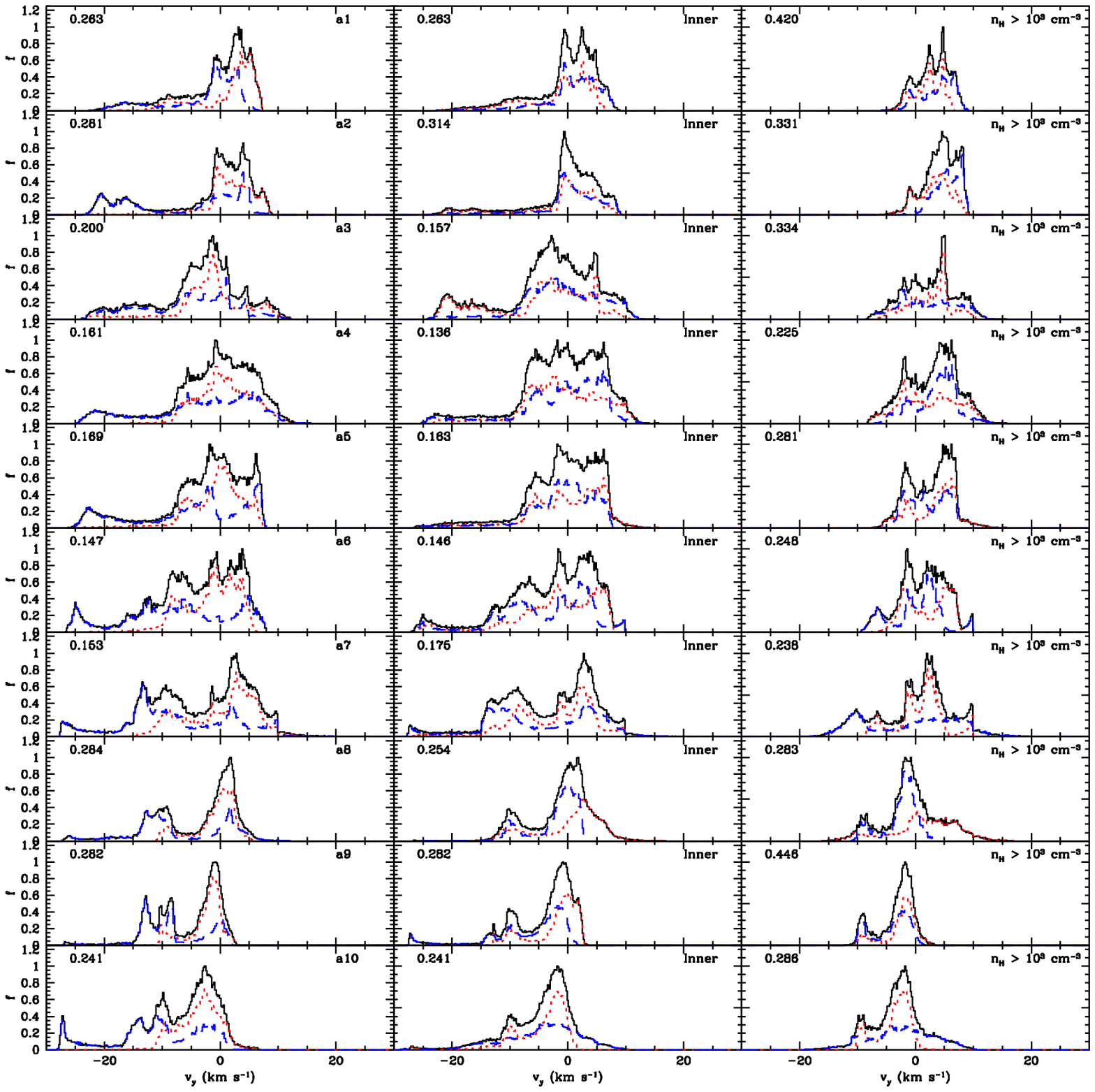}
\end{array}$
\end{center}
\caption{Line of sight velocity spectra for each filament strip (1 to 10, from
top to bottom) for Filament $a$. Left, middle and right columns show
the outer, inner and dense filaments, respectively (see text for
definitions). Total spectra are shown with black solid lines. The
filament strips are also divided into near (red dotted) and far (blue
dashed) side regions with
respect to each strip's center of mass.  
The normalization factor is
given in the top left corner of each panel in units of
${\rm g\:cm^{-2}}$/(km~s$^{-1}$).}\label{fig:vspeca}
\end{figure}

\begin{figure}
\begin{center}$
\begin{array}{c} 
\includegraphics[width=6.0in]{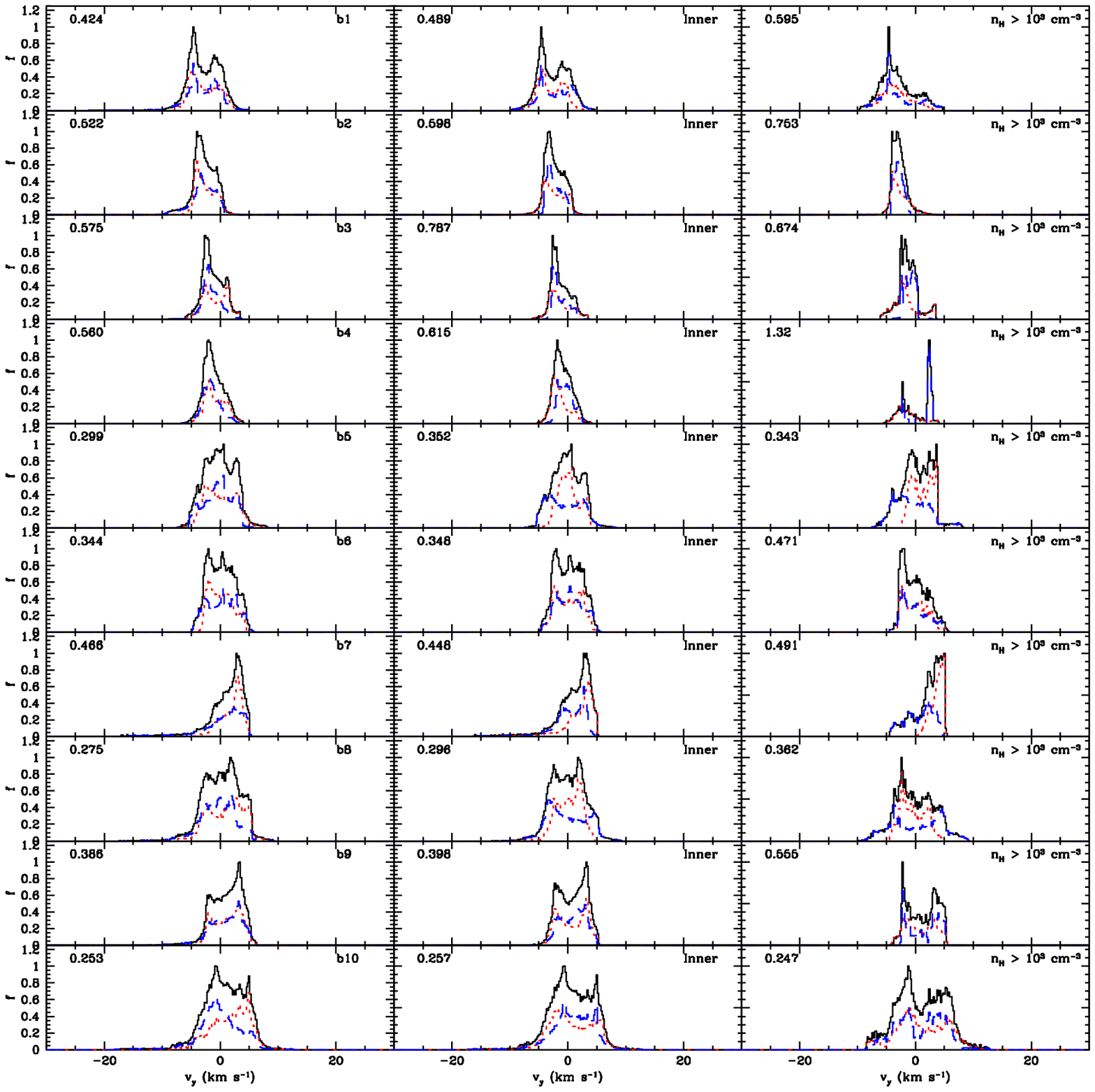}
\end{array}$
\end{center}
\caption{
Same as Figure~\ref{fig:vspeca}, but now for Filament $b$.
}\label{fig:vspecb}
\end{figure}

\begin{figure}
\begin{center}$
\begin{array}{c} 
\includegraphics[width=6.0in]{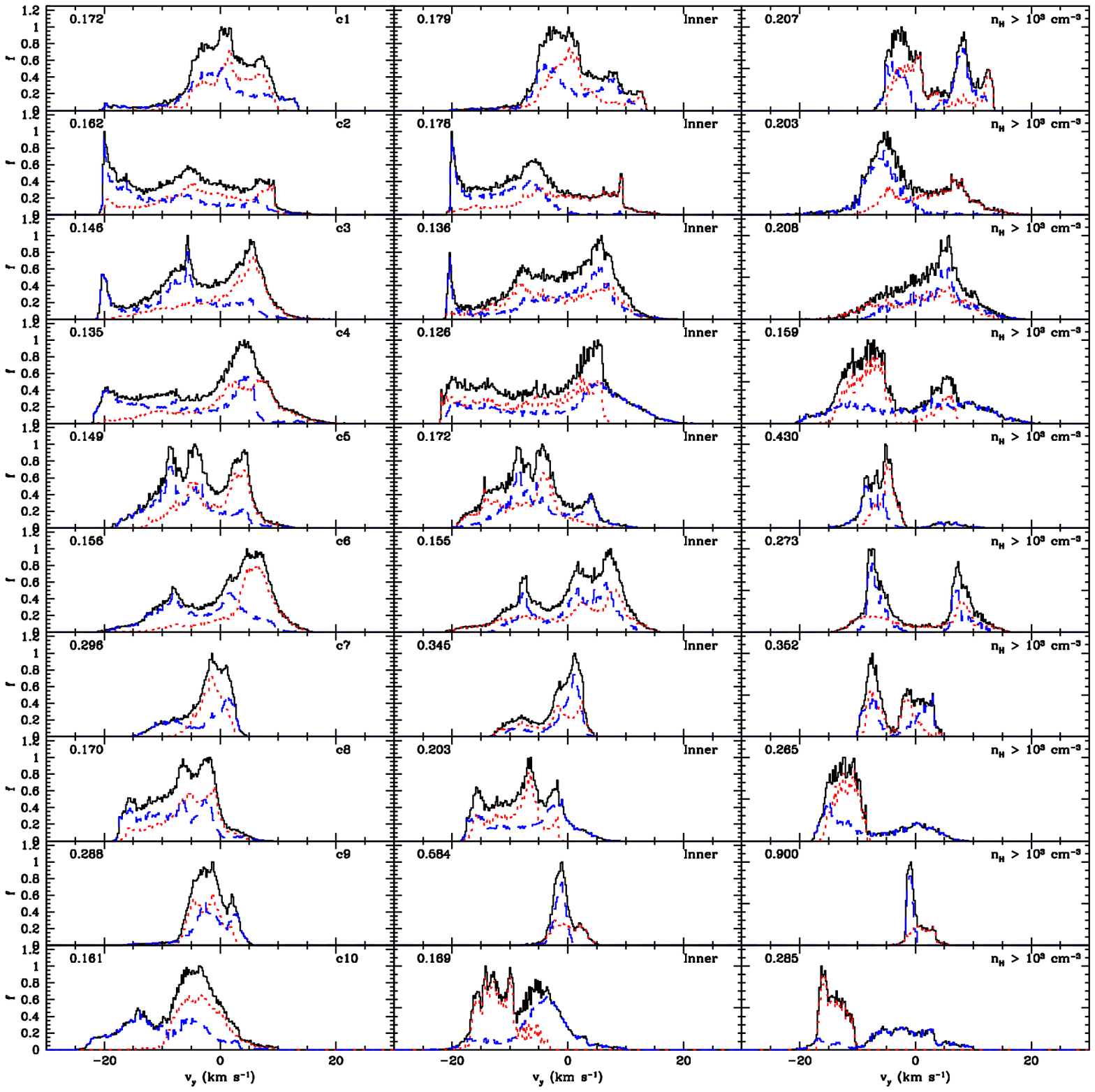}
\end{array}$
\end{center}
\caption{
Same as Figure~\ref{fig:vspeca}, but now for Filament $c$.
}\label{fig:vspecc}
\end{figure}

\begin{figure}
\begin{center}$
\begin{array}{c} 
\includegraphics[width=6.0in]{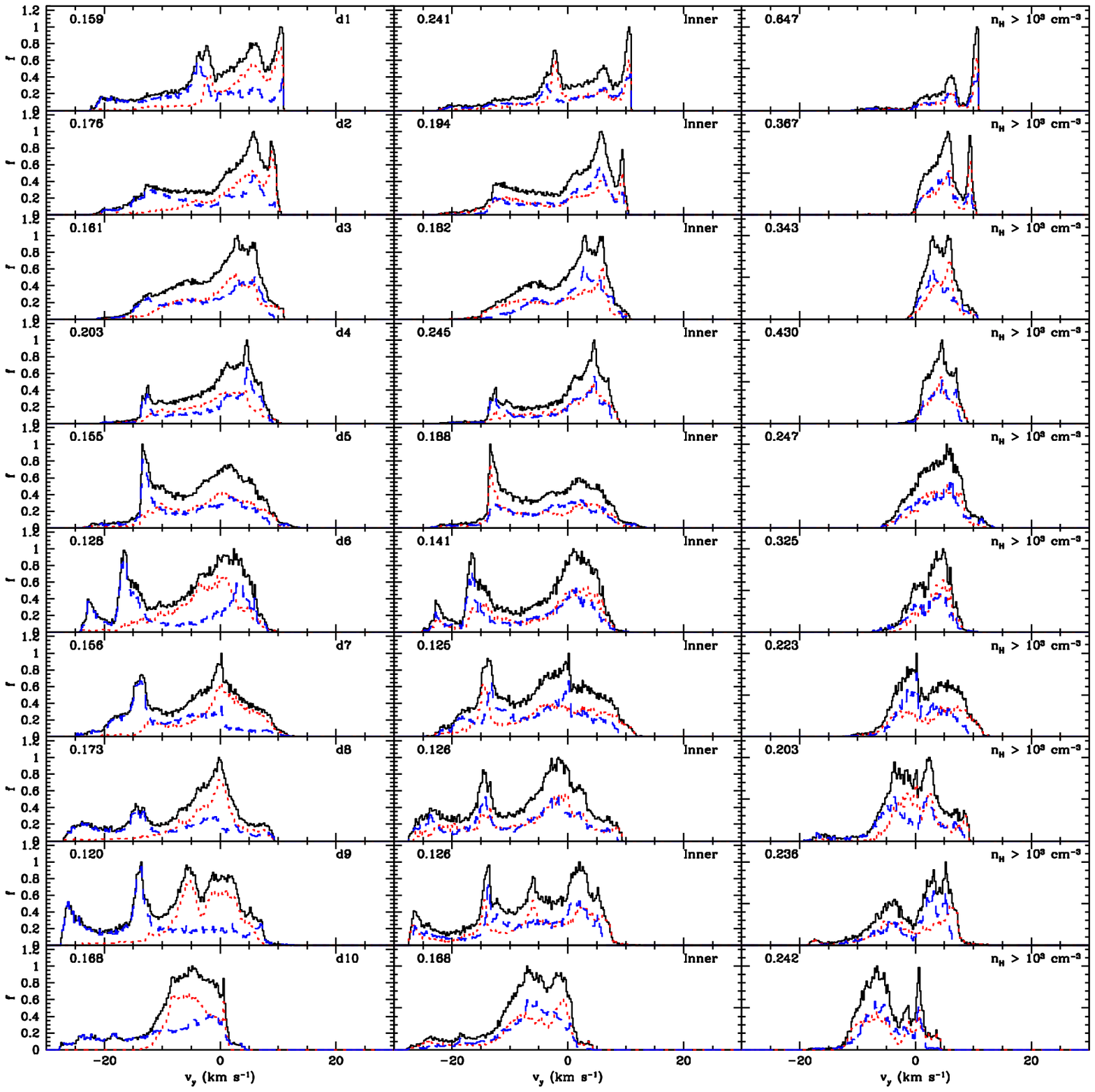}
\end{array}$
\end{center}
\caption{
Same as Figure~\ref{fig:vspeca}, but now for Filament $d$.
}\label{fig:vspecd}
\end{figure}

The third row shows the near side of the outer filament strips in red
and the far side in blue. Thus if we were seeing infall to the
filament, we would expect the near side, red material to be at larger,
red-shifted velocities than the far side, blue material. This kinematic feature is
seen in some regions of all the filaments, although is perhaps less
evident in $b$. To more clearly illustrate such effects, in
Figures~\ref{fig:vspeca}, \ref{fig:vspecb}, \ref{fig:vspecc} and
\ref{fig:vspecd} we also show the velocity histograms (equivalent to
optically thin spectra of an idealized tracer of material at all
densities) of the outer, inner and dense filaments in the 10 strips,
also separating near and far-side material. The disordered kinematics,
i.e., different velocity distributions between near and far sides, are
evident in many cases. However, caution is needed when trying to
interpret blue-shifted far side material as signature of diffuse
infall onto a filament, as such signatures can also arise from
discrete clumps (that may also be infalling).

Finally, in Figure~\ref{fig:posvel} we also separately show the
material at $n_{\rm H} > 10^2, 10^3, 10^4, 10^{5}\:{\rm cm^{-3}}$,
i.e., eventually isolating the clumps that contain the 
dense material that is forming or will form stars. A striking feature
of many of these clumps is the large velocity dispersions, which can
be significantly greater than 10~km~s$^{-1}$. The dynamical properties of the
clumps are discussed below in \S\ref{S:dynamics}.

As mentioned, the line of sight velocity distributions (i.e.,
optically thin spectra) are calculated and shown in
Figures~\ref{fig:vspeca}, \ref{fig:vspecb}, \ref{fig:vspecc} and
\ref{fig:vspecd} for the individual strips of filaments $a$ to
$d$. The spectra for outer, inner and dense filaments are shown,
including separation of near and far sides (for the dense filament
case, the same boundaries for near and far are used as in the inner
filament case, but only showing gas above $n_{\rm H} > 10^{3}\:{\rm
  cm}^{-3}$).

We utilize these spectra to evaluate the mass-weighted mean line of
sight velocities in the strips. These values for the 10 strips in each
filament are then used to measure the best-fit global (50-pc scale)
velocity gradient, weighting the data point from each strip equally
(see Table~\ref{tab:gradients}). These values are $\sim 0.1$ to
$0.2\:{\rm km\:s^{-1}\:pc^{-1}}$, and do not vary much going from
outer to inner to dense filament structures.
On the 5-pc scales from strip to strip centers we also have 9
measurements of velocity gradients, which have averages of
0.446$\pm$0.134, 0.135$\pm$0.0923, 0.553$\pm$0.446,
0.358$\pm$0.312~$\rm km\:s^{-1}\:pc^{-1}$ for the outer filaments
$a$--$d$, respectively (reported in the 10th row for each filament in
Table~\ref{tab:gradients}). Here the uncertainty measurement indicates
the dispersion in the values. Similar results are seen for the inner
and dense filament cases.
The above results indicate that on scales $\sim$~5~pc (similar to
the clump to clump separation scales) the velocity gradients are
several times larger than when averaged over 50-pc scales.



The observed Galactic $\sim100$-pc-scale filaments have global
velocity gradients that are very small. Jackson et al. (2010) find
${\rm d} v/{\rm d} l <0.09\:{\rm km\:s^{-1}\:pc^{-1}}$ in ``Nessie'',
Battersby \& Bally (2012) find ${\rm d} v/{\rm d} r <0.05\:{\rm
  km\:s^{-1}\:pc^{-1}}$ in their 80-pc long cloud, and Ragan et
al. (2014) measure ${\rm d} v/{\rm d} r \simeq 0.06\:{\rm
  km\:s^{-1}\:pc^{-1}}$ as an average of the 7 filaments in their
sample. On smaller scales, there have been some reported measurements
of velocity gradients within IRDCs. For example, Henshaw et al. (2014)
find global velocity gradients of $0.08, 0.07, 0.30\:{\rm
  km\:s^{-1}\:pc^{-1}}$ in several sub-filaments in IRDC H measured on
$\sim$2 parsec scales based on centroid velocities of the dense gas
tracer $\rm N_2H^+(1-0)$. Larger local gradients $\sim 1.5-2.5\:{\rm
  km\:s^{-1}\:pc^{-1}}$ can be present on sub-parsec scales.

The simulated filament $b$ comes closest to matching the above
observed values, while $a$, $c$ and $d$ have 50-pc scale gradients
that are several times larger. More detailed comparison, e.g., of
gradients on smaller $\lesssim 10$~pc scales is warranted, but this
initial study indicates that our simulated filaments have much more
disturbed kinematics on large $\gtrsim 10$~pc scales than the observed
Galactic long filaments.

\begin{table}

\tiny
\begin{center}
\caption{Filament Velocity Gradients} 
\begin{tabular}{c c c}
Filament &   $\bar{v}_{\rm los}$         & ${\rm d} v_{\rm los}/ {\rm d} l$               \\
  &  (km\ s$^{-1}$)  & (km\ s$^{-1}$\ pc$^{-1}$)                     \\
\tableline
$a$ &  -0.223, 0.242, 0.531        &   0.224, 0.223 0.228                           \\
\tableline
$a1$ &  2.85, 2.92, 3.11           &   0.140, 0.210, 0.220          \\
$a2$ &  3.55, 3.97, 4.22           &   0.206, 0.204, 0.148          \\
$a3$ &  2.52, 2.95, 3.48           &   0.278, 0.334, 0.296          \\
$a4$ &  3.91, 4.62, 4.96           &   0.688, 0.618, 0.384          \\
$a5$ &  0.471, 1.53, 1.98          &   0.384, 0.514, 0.470          \\
$a6$ &  -1.45, -1.04, -0.371       &   0.678, 0.616, 0.526         \\
$a7$ &  1.94, 2.04, 2.26           &   1.30, 1.16, 1.20           \\
$a8$ &  -4.57, -3.74, -3.75        &   0.266, 0.404, 0.396          \\
$a9$ &  -5.90, -5.76, -5.73        &   0.0700, 0.138, 0.176          \\
$a10$ & -5.55, -5.07, -4.85        &   [0.446, 0.466, 0.448]$^a$                       \\
\tableline
$b$ &   -0.138, -0.237, -0.292        &  0.0868, 0.0804, 0.0810                   \\
\tableline
$b1$ &     -2.43, -2.31, -2.29       &  0.0460, 0.0100, 0.0280                    \\
$b2$ &     -2.20, -2.36, -2.43       &  0.313, 0.242, 0.270                       \\
$b3$ &     -0.636, -1.15, -1.08      &  0.0492, 0.0810, 0.0840                    \\
$b4$ &     -0.390, -0.745, -0.660    &  0.0474, 0.169, 0.106                                        \\
$b5$ &     -0.153, 0.102, -0.130     &  0.153, 0.0988, 0.151                                        \\
$b6$ &     0.612, 0.596, 0.624       &  0.216,  0.147, 0.159                                     \\
$b7$ &     1.69, 1.33, 1.42          &  0.183,  0.187, 0.202                                     \\
$b8$ &     0.773, 0.396, 0.411       &  0.137,  0.0580, 0.0640                                      \\
$b9$ &     1.46, 0.686, 0.731        &  0.0680, 0.0808, 0.0678                                        \\
$b10$ &    1.12, 1.09,  1.07         &   [0.135, 0.119, 0.126]$^a$                                      \\
\tableline                          
$c$  &      -1.16, -0.584, -2.93    &   0.224, 0.208, 0.210                                    \\
\tableline                        
$c1$ &       4.56, 5.41, 2.90       &   0.664, 0.764, 0.770                                   \\
$c2$ &       1.24, 1.59, -0.952     &   0.390, 0.356, 0.452                                    \\
$c3$ &       3.19, 3.37, 1.31       &   0.344, 0.360, 0.512                                    \\
$c4$ &       1.47, 1.57, -1.25      &   1.13, 1.16, 0.748                                    \\
$c5$ &       -4.19, -4.22, -4.99    &   1.09, 1.15, 0.708                                    \\
$c6$ &       1.27, 1.52, -1.45      &   1.04, 0.912, 0.704                                   \\
$c7$ &       -3.95, -3.04, -4.97    &   0.0480, 0.232, 0.558                                   \\
$c8$ &       -4.19, -4.20, -7.76    &   0.256, 0.400, 0.754                                    \\
$c9$ &       -5.47, -2.20, -3.99    &   0.00600, 0.688, 0.828                                     \\
$c10$ &      -5.50, -5.64, -8.13    &   [0.553, 0.669, 0.671]$^a$                                      \\

\tableline
$d$  &         1.09, 0.970, 2.20        &   0.172, 0.198, 0.221                \\
\tableline
$d1$ &         3.83, 3.97, 6.21         &   0.0280, 0.0260, 0.150                                  \\
$d2$ &         3.97, 4.10, 5.46         &   0.0900, 0.0740, 0.162                                 \\
$d3$ &         3.52, 3.73, 4.65         &   0.188, 0.152, 0.0840                                 \\
$d4$ &         2.58, 2.97, 4.23         &   0.0540, 0.00200, 0.140                                 \\
$d5$ &         2.85, 2.96, 3.53         &   0.549, 0.630, 0.224                                 \\
$d6$ &         0.106, -0.191, 2.41      &   0.263, 0.336, 0.0120                                 \\
$d7$  &        1.42, 1.49, 2.35         &   0.764, 0.856, 0.522                                \\
$d8$  &        -2.40, -2.79, -0.260     &   0.419, 0.511, 0.192                                 \\
$d9$  &        -0.306, -0.235, -1.22    &   0.869, 1.21, 0.824                                   \\
$d10$ &        -4.65, -6.30, -5.34      &   [0.358, 0.422, 0.257]$^a$                                    \\

\end{tabular}

\label{tab:gradients}
\end{center}
$^a$ Average of the 9 strip to strip values.
\end{table}
        
\clearpage
\newpage

\begin{figure}
\begin{center}$
\begin{array}{c} 
\includegraphics[width=6.0in]{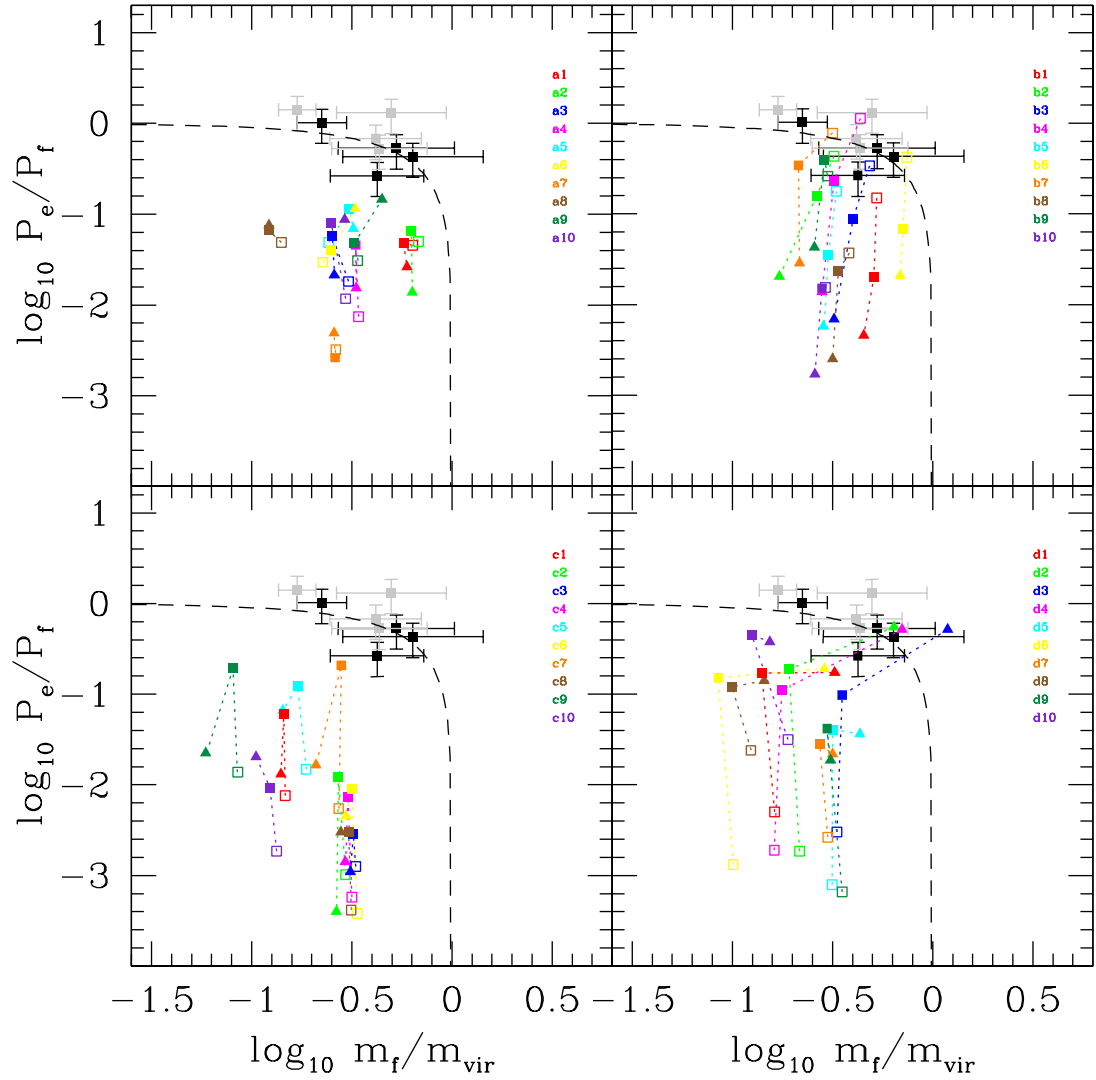} 
\end{array}$
\end{center}
\caption{
Filamentary virial analysis for filament $a$ (top left), $b$ (top
right), $c$ (bottom left) and $d$ (bottom right). Each panel shows the
ratio of envelope, $P_e$, and filament, $P_f$, pressures versus the
ratio of filament mass per unit length, $m_f$, to the virial mass per
unit length, $m_{\rm vir}$, for outer (open squares), inner (solid
squares) and dense (solid triangles) filament regions.  The case for no
magnetic field support is shown as a dashed line (Fiege \& Pudritz
2000).  Points from IRDC H (Hernandez et al. 2012) are plotted as gray (inner filament) and black (outer filament)
points with error bars.
}\label{fig:fiegepudritz}
\end{figure}
\newpage
\clearpage 
\begin{sidewaystable*}[counterclockwise]
\setlength\arraycolsep{3pt}

\tiny
\begin{center}
\caption{Clump Sample} 
\setlength{\tabcolsep}{0.035in} 
\begin{tabular}{c c c c c c c c c c c c}
\tableline
\tableline
Clump & Strip & $x_{c}, y_{c}, z_{c}$  &  Mass & Radius  & $\bar{v}_{\rm x}$ & $\bar{v}_{\rm y}$ & $\bar{v}_{\rm z}$ &  $\bar{\sigma}_{v_{x}}$ &   $\bar{\sigma}_{v_{y}}$ &   $\bar{\sigma}_{v_{z}}$  & $\alpha_{\rm vir}$ \\
    &  &   (pc)                      &   (10$^4$ $M_\odot$) &  (pc)   &  (km\ s$^{-1}$)  &  (km\ s$^{-1}$) &  (km\ s$^{-1}$) &  (km\ s$^{-1}$) &  (km\ s$^{-1}$) &  (km\ s$^{-1}$) &   \\
\tableline
$a_A$ & $a3$      &   469, 541, 2.56  &   0.101       &  0.325   &   0.864     &   1.29             &    0.893         &    1.63            &    2.09        &     1.76        &     1.63                  \\      
$a_{B}$   & $a4$    &   473, 541, 2.53  &   1.03        &  0.485   &   $-$4.55   &   6.24             &    $-$2.06       &    4.41            &    5.98        &     5.52        &     1.95                  \\  
$a_{C}$   & $a6$    &   483, 544, 0.553 &   0.0651      &  0.299   &   4.25      &   $-$2.18          &     1.01         &    1.14            &    1.21        &     0.542       &     0.784                \\
$a_{D}$   & $a7$    &   485, 542, 2.21  &   0.0645      &  0.311   &   5.63      &   2.04             &    0.525         &    1.23            &    1.10        &     1.35        &     0.675                 \\  
$a_{E}$   & $a7$    &   490, 543, 1.36  &   4.90        &  0.713   &   1.33      &   2.45             &   $-$0.660       &    5.43            &    11.0        &     11.1        &     2.03                 \\  
$a_{F}$   & $a8$    &   494, 539, 3.26  &   0.0681      &  0.298   &   $-$0.952  &   0.0854           &    1.50          &    1.30            &    1.58        &     1.63        &     1.27                  \\  
$a_{G}$   &  $a10$   &   502, 539, 2.25  &   0.181       &  0.336   &   9.00      &  $-$1.54           &   1.52           &    1.93            &    2.86        &     2.63        &     1.77                  \\  
\tableline
$b_{A}$   & $b1$    &  624, 586, 2.03     &   0.316       &  0.444   &   $-$2.21   &  $-$6.85  &   $-$0.0531 & 3.14        &   1.76     &  2.57      &  1.61   \\
$b_{B}$   & $b1$    &  626, 590, 2.43     &   0.621       &  0.485   &   $-$2.15   &  $-$6.75  &   0.0253    & 4.50        &   2.59     &  4.23      &  1.84   \\
$b_{C}$   & $b3$   &  628, 597, 1.99     &   0.239       &  0.391   &   $-$1.00   &  $-$4.10  &   0.185     & 2.45        &   1.75     &  3.12      &  1.15   \\
$b_{D}$   & $b4$    &  628, 601, 1.89     &   0.189       & 0.438    &   $-$0.494  &  $-$3.60  &   $-$0.320  & 2.21        &   1.05     &  2.10      &  1.31   \\
$b_{E}$   & $b5$    &  629, 608, 1.51     &   0.622       & 0.502    &   0.504     &  0.996    &   $-$0.0556 & 4.48        &   1.92     &  4.41      &  1.88   \\
$b_{F}$   & $b6$    &  630, 612, 0.798    &   0.151       & 0.362    &   1.13      &  1.70     &   1.48      & 2.45        &   1.25     &  2.06      &  1.67   \\
$b_{G}$   & $b6$    &  630, 614, 0.929    &   0.200       & 0.420    &   1.41      &  $-$1.23  &   0.758     & 2.31        &   1.17     &  1.62      &  1.30   \\
$b_{H}$   & $b8$    &  629, 620, 0.628    &   0.0972      &  0.326   &   $-$0.399  &  4.12     &   $-$0.283  & 1.94        &   0.945    &   1.85     &  1.47   \\
$b_{I}$   & $b8$    &  629, 624, 0.281    &   0.993       &  0.558   &   0.734     &  4.71     &   $-$0.0817 & 5.52        &   1.89     &  4.94      &  1.99   \\
$b_{J}$   & $b9$    &  629, 627, 0.617    &   0.141       &  0.379   &   0.451     &  0.629    &   $-$0.538  & 2.20        &   1.20     &  1.97      &  1.51   \\
$b_{K}$   &  $b10$   &  628, 632, $-$0.771 &   1.49        &  0.580   &   1.49      &  4.40     &   0.261     & 5.97        &   2.24     &  6.85      &  1.62   \\
\tableline
$c_{A}$    &  $c2$  &  631, 233, 15.2     &   4.94        &  0.762   &  2.50       &  2.69     & 1.16        & 8.53        &   10.3     &  2.68      &  1.30   \\
$c_{B}$    &  $c3$  &  624, 237, 17.3     &   4.09        &  0.723   &  3.03       &  0.665    & $-$0.386    & 9.29        &   7.94     &  2.65      &  1.77   \\
$c_{C}$    &  $c3$  &  626, 238, 17.5     &   0.442       &  0.472   &  5.17       &  $-$7.00  & 2.07        & 4.70        &   1.86     &  2.35      &  2.74   \\
$c_{D}$    &  $c4$  &  624, 243, 17.0     &   0.203       &  0.347   &  6.30       &  1.68     & 3.22        & 1.88        &   2.46     &  2.67      &  0.705   \\
$c_{E}$    &  $c4$  &  621, 245, 18.0     &   0.0980      &  0.329   &  4.64       &  $-$3.91  & 7.75        & 1.74        &   1.22     &  1.47      &  1.18   \\
$c_{F}$    &  $c4$  &  618, 245, 18.9     &   5.68        &  0.781   &  1.57       &  0.294    & $-$0.0265   & 11.8        &   3.47     &  9.62      &  2.23   \\
$c_{G}$    &  $c6$  &  612, 252, 20.9     &   4.03        &  0.732   &  1.83       &  $-$1.46  & $-$1.30     & 8.98        &   3.99     &  10.3      &  1.70   \\
$c_{H}$    &  $c6$  &  611, 253, 20.3     &   0.173       &  0.354   &  12.2       &  $-$10.9  & $-$0.966    & 2.25        &   1.69     &  2.67      &  1.21   \\
$c_{I}$    &  $c7$  &  608, 259, 23.3     &   0.102       &  1.05    &  $-$0.883   &  $-$0.102 & 1.82        & 0.939       &   1.75     &  1.75      &  1.33   \\
$c_{J}$    &  $c8$  &  602, 265, 25.4     &   2.62        &  0.708   &  $-$3.24    &  4.18     &  6.82       & 7.56        &   2.24     &  7.07      &  1.80   \\
$c_{K}$    & $c10$   &  593, 276, 28.8     &   0.297       &  0.395   &  $-$3.48    &  $-$3.81  & $-$13.0     & 3.49        &   3.19     &  3.07      &  1.88   \\
\tableline
$d_{A}$   &  $d1$  &  782, 178, 6.87     &   0.0445      &  0.286   &  3.86       &  3.39     & 0.268       & 1.24        &   0.867    &  1.34      &  1.15   \\      
$d_{B}$   &  $d3$  &  784, 187, 5.85     &   0.105       &  0.357   &  5.32       &  4.38     & 0.232       & 1.65        &   1.48     &  1.75      &  1.07   \\  
$d_{C}$   &  $d5$  &  787, 198, 6.54     &   1.16        &  0.533   &  3.95       &  4.38     & 1.09        & 6.06        &   3.58     &  5.87      &  1.96   \\
$d_{D}$   &  $d7$  &  787, 206, 7.12     &   1.76        &  0.575   &  1.89       &  4.69     & 0.0727      & 7.02        &   3.77     &  7.59      &  1.87   \\ 
$d_{E}$   &  $d9$  &  784, 216, 7.87     &   2.03        &  0.600   &  0.152      &  3.70     & 0.186       & 6.92        &   3.82     &  8.44      &  1.64   \\


\tableline
\tableline

\end{tabular}
\label{tab:clumps}
\end{center}
\end{sidewaystable*}
\clearpage
\subsection{Filament and Clump Dynamics}\label{S:dynamics}
The simulated filaments have formed by large-scale collapse of
self-gravitating ``GMCs,'' in which collapse is unable to be resisted
by magnetic fields or local feedback from star formation. Some
resistance to collapse is provided by the turbulent and shearing
motions present in and around the clouds due to their galactic
environment. By 4~Myr after the beginning of the simulation, the four
filaments that we have chosen for analysis are in various stages of
collapse and fragmentation. Here we assess their dynamical state, i.e., how
close are they to virial equilibrium?

We carry out a filamentary virial analysis for each filament strip
following Fiege \& Pudritz (2000), who derived the following equation
satisfied by pressure-confined, nonrotating, self-gravitating,
filamentary (i.e., lengths $\gg$ widths) clouds that are in virial equilibrium:
\begin{equation}
\frac{P_e}{P_f} = 1 - \frac{m_f}{m_{\rm vir}}\left(1-\frac{{\cal M}_l}{|W_l|}\right).
\label{eq:fiege}
\end{equation}
Here, $P_e$ is the external envelope pressure at the filament surface,
$P_f=\rho_f \sigma_f^2$ is the average total pressure in the filament, $m_f$
is the filament mass per unit length, $m_{\rm vir} \equiv 2 \sigma_f^2/G$ is the
filament virial mass per unit length, ${\cal M}_f$ is the magnetic energy per
unit length
and $W_f=-m_f^2G$ is the gravitational energy per unit length.

For each filament strip, for the cases of outer, inner and dense
filament regions, we measure the mass-weighted 1D velocity dispersion
in the direction of observation that is in the galactic plane and
orthogonal to the filament main axis. These values are listed in
Table~\ref{tab:filamentkinematics}, along with the dispersions
measured in other directions. The velocity dispersions are supersonic,
with Mach numbers of about 20 for gas that is cooled to about
10~K. Note, that the z-direction velocity dispersions are similar in
size to those in directions in the Galactic plane (unlike GMC
motions), indicating that approximately isotropic support may be
possible from these turbulent motions.

We also measure the velocity dispersions in the surrounding envelope
regions, which, together with the density of these regions allows us
to assess the turbulent pressure that acts as a surface term, $P_{e} =
\rho_e\sigma_e^2$, affecting the virial equilibrium of the
filaments. Here we choose to measure the volume-averaged density in
the envelope region and the mass-weighted velocity dispersion, which
is comparable to the quantities derived observationally for density
(from extinction maps) and velocity dispersion from $^{13}$CO and
C$^{18}$O spectra (e.g., Hernandez \& Tan 2011; Hernandez et
al. 2012). We discuss, below, the effects of these choices on our
results.

In Figure~\ref{fig:fiegepudritz}, we compare the simulated filament
strips with the non-magnetic (${\cal M}_f=0$) form of
equation~\ref{eq:fiege}. For each strip we show the outer, inner and
dense filament results connected by a line. In general, the filaments
have $0.1\gtrsim m_f/m_{\rm vir}\lesssim 1$, but this is not typically
due to higher envelope pressure. Thus, most filament strip regions
appear to not yet be virialized, having very disordered kinematics due
to infall motions and motions associated with dense, spheroidal clumps
that have already fragmented from the filament, discussed below. Only
in a few regions of filaments $b$ (outer) and $d$ (dense) are
conditions closer to filamentary virial equilbrium. Note, filament
$b$'s inner region is highly fragmented into clumps and so when
defining the envelope material around the ``dense'' filament material,
we derive relatively low values leading to values of $P_e/P_f$ much
lower than expected by filamentary virial equilbrium. Filament $d$,
being relatively less fragmented into clumps, shows inner regions
conditions that are closer to virial equilibrium.


Here we examine more quantitatively how fragmentation of filaments
into clumps that are themselves approximately virialized can lead to
small values of $m_f/m_{\rm vir}$ for the filament.  Consider the case
when a fraction, $\epsilon_{\rm cl}$, of the filament strip mass, $M$,
has condensed into a virialized clump of radius $R_{\rm cl}$ and
velocity dispersion $\sigma_{\rm cl}$, which is given by $\sigma_{\rm
  cl}^2 = G\epsilon_{\rm cl}M/(5R_{\rm cl})$ (note, corrections for
ellipticity and central concentration of the clump are typically
modest, Bertoldi \& McKee (1992)). The velocity dispersion expected from the filamentary
virial theorem is given by $\sigma_f^2 = G m_f/2 = G M/(2L)$, where
$L$ is the length of the strip, i.e., 5~pc in the cases considered
here. Then,
\begin{equation}
\frac{\sigma_{\rm cl}^2}{\sigma_f^2} = \frac{2L}{5R_{\rm cl}} \epsilon_{\rm cl},
\end{equation}
and so with $\sigma_{\rm tot}^2 = \epsilon_{\rm cl} \sigma_{\rm cl}^2
+ (1-\epsilon_{\rm cl})\sigma_f^2$, then
\begin{equation}
\frac{\sigma_{\rm tot}^2}{\sigma_f^2} 
= (1- \epsilon_{\rm cl}) + \frac{2}{5}\frac{L}{R_{\rm cl}}\epsilon_{\rm cl}^2
\label{eq:clumpboost}
\end{equation}
If $\epsilon_{\rm cl}\gtrsim 5R_{\rm cl}/(2L)$, then the total velocity
dispersion begins to increase above that expected from the filamentary
virial theorem. We will see below that identified clumps have typical
radii of $\sim 0.5$~pc, so then the critical efficiency for increasing
velocity dispersion is $\simeq 0.25 (R_{\rm cl}/0.5\:{\rm pc})$.


For example, strip $a7$, which has a relatively small value of
$m_f/m_{\rm vir}\simeq 0.25$, has a $5\times 10^4\:M_\odot$ clump, so
that $\epsilon_{\rm cl}\simeq 0.8$. It has a radus of about $0.7$~pc
and a mean velocity dispersion in the $y$ direction of $11\:{\rm
  km\:s^{-1}}$ (it has a virial parameter of $\alpha_{\rm
  vir}=$2.0). The total velocity dispersion for the $a7$ strip is
$10.2\:{\rm km\:s^{-1}}$. However, with its mass per unit length of
$1.28\times 10^4\:M_\odot\:{\rm pc}^{-1}$, one would expect a
virialized velocity dispersion of only 5.2~${\rm km\:s^{-1}}$. This
difference is explained by the clump-boosting factor for velocity
dispersion squared predicted by equation~\ref{eq:clumpboost} of about
2.0, i.e., a factor of 1.4 for the velocity dispersion. Since the
clump itself is moderately super-virial, then this explains why the
actual velocity dispersion increase is larger. So when placing the
$a7$ strip on the Fiege-Pudritz diagram, we expect $m_l/m_{\rm vir}$
to be $\simeq 0.25$. Similarly, the dominance within the filament
strip of spheroidal clump dynamics causes a very low value of
$P_e/P_f$.





Given the difficulties of resolving clumps and their fragmentation
from the filament, we have focused mostly on filament dynamics for
comparison of simulation results with observations. However, for
completeness, here we give a brief assessment of the dynamical state
of the identified clumps. For each of the clumps identified by the
threshold density of $10^{5}\:{\rm cm^{-3}}$, we measure the mass in
the cells, the mass-weighted velocity dispersion about the center of
mass velocity, and the mean half-mass radius, $R_{1/2}$, in each of
the $x$, $y$ and $z$ directions, and the average value. We then
evaluate the virial parameter, $\alpha_{\rm vir} = 5 \sigma^2 R_{1/2}/
(G M_{1/2})$ (Bertoldi \& McKee (1992)) at this half-mass scale. The results are
summarized in Table~\ref{tab:clumps}.

The clumps tend to have virial parameters $\simeq 1$, and almost
always $<2$, indicating they are gravitationally bound and perhaps
moderately supervirial (although a contribution from surface pressure
would also raise the virialized value of $\alpha_{\rm vir}$ above
unity). We tentatively conclude, with the caveat that increased numerical
resolution is needed, that these clump structures are much closer to
virial equilibrium than their parental filaments, which is to be
expected given their much shorter dynamical times.

\section{Discussion and Conclusions}\label{S:conclusion}

We have continued the study begun in Van Loo et al. (2013), following
the evolution of a kiloparsec-scale patch of a galactic disk extracted
from a global disk simulation down to $\sim 0.1\ \rm pc$
resolution. We have followed collapse for 4~Myr, which is ample time
for the formation of dense filaments and clumps from the initial GMCs
and, in the run allowing star formation, for significant star
formation activity.

The main goal has been to study the detailed structural, kinematic and
dynamical properties of filamentary clouds that are in relatively
early stages of collapse at $t=$~4~Myr. Even in the run where star
formation is allowed (above the threshold density of $n_{\rm
  H}=10^6\:{\rm cm^{-3}}$), almost no star formation has yet
occurred in our sample of filaments. These properties have been compared to those of observed
filamentary IRDCs, which are also thought to be in a relatively early
stage of collapse and star formation. Note, IRDCs are thus thought to
be relatively unaffected by internal feedback from star formation,
simplifying the comparison of simulation with observation by avoiding
having to simulate star formation feedback, which requires much high
resolution or more uncertain sub-grid models.

Our main conclusion is that the simulated filaments, which are forming
from global collapse of gravitationally unstable GMCs mediated by
galactic shear driven turbulence, show significant differences from
observed IRDCs. The filaments and their surrounding 50-pc scale
regions often have dense gas mass fractions, e.g., at $\Sigma>1\:{\rm
  g\:cm^{-2}}$ that are larger than even the most extreme IRDCs. The
simulated filaments show large dispersions in their mass per unit
lengths, caused by their fragmentation into dense clumps. The
simulated filaments have more disordered kinematics, including
velocity gradients as measured on 50 and 5 pc length scales. These
more disordered kinematics equate to larger velocity dispersions than
expected of virialized filaments.

The implications of these results are that IRDCs do not form by fast
global collapse of gravitationally unstable GMCs. Mediation,
regulation and slowing of collapse by dynamically strong magnetic
fields seems to be the most promising mechanism by which to reconcile
simulations with observed IRDCs. This scenario is given support by the
recent observational results of Pillai et al. (2015), who infer
$\sim$~mG magnetic fields and sub-Alfv\'enic turbulence in IRDCs
G11.11-0.12 and G0.253+0.016 from the ordered orientations of sub-mm
dust emission polariazation vectors.  
Lower (0.5~pc) resolution simulations of the same initial conditions
as our study, but including magnetic fields, have been recently
presented by Van Loo et al. (2015, Paper II). A future goal is to
extend these to 0.1~pc or higher resolution to be able to examine the
effect of magnetic fields on filament structure, kinematics and
dynamics.

A more general output of this paper has been the presentation of a
range of metrics of cloud, especially filamentary cloud, properties
related to structure, kinematics and dynamics. These properties are
presented for multi-phase ISM, especially molecular, clouds evolving
under pure self-gravitating hydrodynamics, i.e., without inclusion of
magnetic fields or star formation feedback (although with the focus on
nearly starless clouds, this feedback is expected to be
limited). These cloud metrics include structural properties on 25 and
50~pc region scales, including PDFs of mass surface density and the
fraction of gas above $\rm 1\:g\:cm^{-2}$. Structural properties of
filaments include mass per unit length, dispersion in mass per unit
length, filament and envelope densities and lateral widths. Kinematic
properties include filament and envelope velocity dispersions,
comparison of mass per unit lengths with virial mass per unit lengths,
external to internal pressures, and velocity gradients on various
scales. Standard kinematic and dynamical properties of clumps, forming
in the filaments, have also been presented. These properties can all
be compared to observed clouds, especially those of IRDCs, as well as
future simulations that include additional physics, especially
magnetic fields and star formation feedback.

\acknowledgements We thank Paola Caselli and Sam Falle for useful
discussions, and Elizabeth Tasker for providing the initial conditions of our simulations.  M.J.B. acknowledges support from the STARFORM Sinergia project funded by the Swiss National Science Foundation.  S.v.L. acknowledges support from the
Theory Postdoctoral Fellowship from UF Department of Astronomy and
College of Liberal Arts and Sciences and from the SMA Postdoctoral
Fellowship of the Smithsonian Astrophysical
Observatory. J.C.T. acknowledges support from NSF CAREER grant
AST-0645412; NASA Astrophysics Theory and Fundamental Physics grant
ATP09-0094; NASA Astrophysics Data Analysis Program
ADAP10-0110. Resources supporting this work were provided by the
University of Florida Research Computing and the NASA High-End
Computing (HEC) Program through the NASA Advanced Supercomputing (NAS)
Division at Ames Research Center.

\end{document}